\documentclass[12pt,letterpaper]{article}

\usepackage{xcolor, colortbl}
\usepackage{amssymb, amsmath, amsthm}
\usepackage{graphicx}
\usepackage{natbib}
\usepackage{float}
\usepackage{setspace}
\usepackage{algorithm}
\usepackage{algorithmic}

\definecolor{lightgray}{gray}{0.9}
\newtheorem{lemma}{Lemma}
\newtheorem{proposition}{Proposition}
\newtheorem{corollary}{Corollary}
\newcommand{\di}{\text{d}}

\newcommand{\bA}{\mathbf{A}}
\newcommand{\bB}{\mathbf{B}}
\newcommand{\bc}{\mathbf{c}}
\newcommand{\bC}{\mathbf{C}}

\newcommand{\bF}{\mathbf{F}}
\newcommand{\bK}{\mathbf{K}}
\newcommand{\bL}{\mathbf{L}}
\newcommand{\bm}{\mathbf{m}}
\newcommand{\bz}{\mathbf{z}}

\newcommand{\bR}{\mathbf{R}}
\newcommand{\bS}{\mathbf{S}}

\newcommand{\bP}{\mathbf{P}}
\newcommand{\bw}{\mathbf{w}}
\newcommand{\bu}{\mathbf{u}}

\newcommand{\bV}{\mathbf{V}}
\newcommand{\bW}{\mathbf{W}}
\newcommand{\bT}{\mathbf{T}}
\newcommand{\bX}{\mathbf{X}}
\newcommand{\bx}{\mathbf{x}}
\newcommand{\by}{\mathbf{y}}

\renewcommand{\epsilon}{\varepsilon}
\renewcommand{\hat}{\widehat}
\renewcommand{\tilde}{\widetilde}
\renewcommand{\leq}{\leqslant}
\renewcommand{\geq}{\geqslant}

\newcommand{\distn}[1]{\mathcal{#1}}

\newcommand{\Em}{\mathbb E}

\newcommand{\gvn}{\,|\,}
\newcommand{\e}{\text{e}}

\newcommand{\vect}[1]{\boldsymbol #1}
\newcommand{\vtheta}{\vect{\theta}}

\newcommand{\vbeta}{\vect{\beta}}

\newcommand{\vepsilon}{\vect{\epsilon}}

\newcommand{\vSigma}{\vect{\Sigma}}

\newcommand{\vpsi}{\vect{\psi}}
\newcommand{\vgamma}{\vect{\gamma}}

\newcommand{\valpha}{\vect{\alpha}}

\newcommand{\vsigma}{\vect{\sigma}}

\DeclareMathOperator{\diag}{diag}
\newcommand{\matlab}{\mathrm{M}\mathrm{{\scriptstyle ATLAB}}}

\bibpunct{(}{)}{;}{a}{,}{,}

\begin{document}


\title{Asymmetric Conjugate Priors for Large \\ Bayesian VARs}

\author{Joshua C.C. Chan\thanks{I would like to thank Todd Clark, Gary Koop, Christian Matthes, Michele Piffer, Co-Editor Tao Zha and two anonymous referees for many constructive comments and useful suggestions that have substantially improved a previous version of the paper.} \\
{\small Purdue University}  
}

\date{October 2021}


\maketitle

\onehalfspacing

\begin{abstract}

\noindent Large Bayesian VARs are now widely used in empirical macroeconomics. One popular shrinkage prior in this setting is the natural conjugate prior as it facilitates posterior simulation and leads to a range of useful analytical results. This is, however, at the expense of modeling flexibility, as it rules out cross-variable shrinkage --- i.e., shrinking coefficients on lags of other variables more aggressively than those on own lags. We develop a prior that has the best of both worlds: it can accommodate cross-variable shrinkage, while maintaining many useful analytical results, such as a closed-form expression of the marginal likelihood. This new prior also leads to fast posterior simulation --- for a BVAR with 100 variables and 4 lags, obtaining 10,000 posterior draws takes less than half a minute on a standard desktop. We demonstrate the usefulness of the new prior via a structural analysis using a 15-variable VAR with sign restrictions to identify 5 structural shocks.


\bigskip

\noindent Keywords: shrinkage prior, marginal likelihood, optimal hyperparameters, structural VAR, sign restrictions

\bigskip

\noindent JEL classifications: C11, C52, C55, E44

\end{abstract}

\thispagestyle{empty}


\newpage

\section{Introduction}

Large Bayesian vector autoregressions (BVARs) have become increasingly popular in empirical macroeconomics for forecasting and structural analysis since the influential work by \citet*{BGR10}. Prominent examples include \citet*{CKM09}, \citet{koop13}, \citet{KK13} and \citet{KP19}. VARs tend to have a lot of parameters, and the key that makes these highly parameterized VARs useful is the introduction of shrinkage priors. For large BVARs, one commonly adopted prior is the natural conjugate prior, which has a few advantages over alternatives. First, this prior is conjugate, and consequently it gives rise to a range of useful analytical results, including a closed-form expression of the marginal likelihood.\footnote{An analytical expression for the marginal likelihood is valuable for many purposes. First, it is useful for model selection --- e.g., choosing the lag length in BVARs. Second, it can be used to select prior hyperparameters that control the degree of shrinkage. Examples include \citet{DS04}, \citet*{SS15} and \citet*{CCM15}. This approach of selecting hyperparameters is incorporated in the BEAR $\matlab$ toolbox developed by the European Central Bank \citep*{DLV16}.} Second, the posterior covariance matrix of the VAR coefficients under this prior has a Kronecker product structure, which can be used to speed up computations. 

On the other hand, a key limitation of the natural conjugate prior is that the prior covariance matrix of the VAR coefficients needs to have a Kronecker product structure, which implies cross-equation restrictions that might not be reasonable. In particular, this Kronecker structure requires symmetric treatment of own lags and lags of other variables. In many applications one might wish to shrink the coefficients on other variables' lags more strongly to zero than those of own lags. This cross-variable shrinkage, however, cannot be implemented using the natural conjugate prior due to this Kronecker structure. \citet{CCM15} summarize this dilemma between computational convenience and prior flexibility as: ``While the pioneering work of \citet{litterman86} suggested it was useful to have cross-variable shrinkage, it has become more common to estimate larger models without cross-variable shrinkage, in order to have a Kronecker structure that speeds up computations and facilitates simulation."

We develop a prior that solves this dilemma --- this new prior allows asymmetric treatment between own lags and lags of other variables, while it maintains many useful analytical results, such as a closed-form expression of the marginal likelihood. In addition, we exploit these analytical results to develop an efficient method to simulate directly from the posterior distribution --- we obtain \emph{independent} posterior draws and avoid Markov chain Monte Carlo (MCMC) methods altogether. For a BVAR with 100 variables and 4 lags, simulating 10,000 posterior draws under this new asymmetric conjugate prior takes less than 30 seconds. 

To develop this asymmetric conjugate prior, we first write the BVAR in a recursive structural form, under which the error covariance matrix is diagonal. We then adopt an equation-by-equation estimation approach in the spirit of \citet*{CCM19}. In particular, we assume that the parameters are \emph{a priori} independent across equations --- i.e., the joint prior density is a product of densities, each for the set of parameters in each equation. Under this setup, we show that if the VAR coefficients and the error variance in each equation follows a normal-inverse-gamma prior, the posterior distribution has the same form --- i.e., it is a product of normal-inverse-gamma densities. It is useful to emphasize that this particular structural-form parameterization is intended to be a computational device and it does not commit the user to a recursive identification scheme. In particular, one can recover the reduced-form parameters from the structural-form parameter, and the posterior sampler can be used in conjunction with other identification schemes, as demonstrated in the application.

To help elicit the hyperparameters in this asymmetric conjugate prior, we prove that if we assume a standard inverse-Wishart prior on the reduced-form error covariance matrix, the implied prior on the structural-form impact matrix and error variances is a product of normal-inverse-gamma densities; and vice versa. Hence, using this proposition, we can first elicit the hyperparameters in the reduced-form prior, which is often more natural, and then obtain the implied hyperparameters in the structural-form prior. Since we can directly specify prior beliefs on the reduced-form error covariance matrix, this proposition implies that the proposed prior --- with carefully chosen hyperparameters --- is independent of the order of the variables.

We demonstrate the usefulness of the proposed asymmetric conjugate prior and the associated posterior sampler by revisiting the empirical study in \citet{FRS19}, who use 6-variable VAR to identify 5 structural shocks using a set of sign restrictions on the contemporaneous impact matrix. Here we augment their system with a set of additional variables and consider a 15-variable VAR, which is one of the largest VARs identified with sign restrictions considered in the literature so far. Even though there are good reasons to consider larger systems --- such as concerns of informational deficiency and non-unique mapping from economic variables to data --- empirical works that impose sign restrictions typically use small or medium VARs (e.g., up to 6 or 7 variables) due to the computational burden. With more variables and sign restrictions, it is clear that conventional approaches that use standard Gibbs samplers are too computationally expensive. In contrast, using the proposed asymmetric conjugate prior and the associated efficient sampler, it is feasible to conduct structural analysis with a large number of sign restrictions, which helps sharpen inference.


The rest of the paper is organized as follows. We first introduce in Section~\ref{s:BVAR} a reparameterization of the reduced-form BVAR and the new asymmetric conjugate prior. We then derive the associated posterior distribution and the marginal likelihood. Section~\ref{s:extensions} discusses a few extensions of the standard BVAR, and outlines the corresponding sampling schemes. It is followed by a structural analysis using sign restrictions to illustrate the usefulness of the proposed prior in Section~\ref{s:application}. Lastly, Section~\ref{s:conclusion} concludes and briefly discusses some future research directions.

\section{Bayesian VARs and Conjugate Priors} \label{s:BVAR}

Let $\by_t= (y_{1,t},\ldots,y_{n,t})'$ be an $n \times 1$ vector of endogenous variables at time $t$. A standard VAR can be written as:
\begin{equation} \label{eq:VAR}
    \by_t = \tilde{\mathbf{b}} + \tilde{\bB}_{1} \by_{t-1} + \cdots + \tilde{\bB}_{p} \by_{t-p} +
		\tilde{\vepsilon}_t^y, \quad \tilde{\vepsilon}_t^y \sim \distn{N}(\mathbf{0}, \tilde{\vSigma}),
\end{equation}
where $\tilde{\mathbf{b}}$ is an $n\times 1 $ vector of intercepts, $\tilde{\bB}_{1}, \ldots, \tilde{\bB}_{p}$ are $n \times n$ VAR coefficient matrices and $\tilde{\vSigma}$ is a full covariance matrix. 

The parameters in this model can be naturally divided into two blocks: the error covariance matrix $\tilde{\vSigma}$ and the matrix of intercepts and VAR coefficients, i.e., $\tilde{\mathbf{B}} = (\tilde{\mathbf{b}}, \tilde{\bB}_{1}, \cdots, \tilde{\bB}_{p})'.$ Under this parameterization, there is a conjugate prior on $(\tilde{\mathbf{B}}, \tilde{\vSigma})$, namely, the normal-inverse-Wishart distribution:
\[
	\tilde{\vSigma}\sim\distn{IW}(\tilde{\nu}_0,\tilde{\bS}_0), \quad (\text{vec}(\tilde{\mathbf{B}})\gvn\tilde{\vSigma})\sim\distn{N}(\text{vec}(\tilde{\mathbf{B}}_0), 
	\tilde{\vSigma}\otimes \tilde{\bV}),
\]
where $\otimes$ denotes the Kronecker product, $\text{vec}(\cdot)$ vectorizes a matrix by stacking the columns from left to right and $\distn{IW}$ denotes the inverse-Wishart distribution. This prior is commonly called the natural conjugate prior and can be traced back to \citet{zellner71}. For textbook treatment of this prior and the associated posterior distribution, see, e.g., \citet{KK10}, \citet{karlsson13} or \citet{chan20b}.

The main advantage of the natural conjugate prior is that it gives rise to a range of analytical results. For example, the associated posterior and one-step-ahead predictive distributions are both known; the marginal likelihood is also available in closed-form. These analytical results are useful for a variety of purposes. For instance, the closed-form expression of the marginal likelihood under the natural conjugate prior can be used to calculate optimal hyperparameters, as is done in \citet{DS04}, \citet*{SS15} and \citet*{CCM15}. The direct sampling algorithm to draw from the posterior distribution of $(\text{vec}(\tilde{\mathbf{B}}), \tilde{\vSigma})$ can be used to develop efficient posterior samplers to estimate large, heteroscedastic Bayesian VARs. Examples include \citet*{CCM16} and \citet{chan20}.

On the other hand, one key drawback of the natural conjugate prior is that the prior covariance matrix of $\text{vec}(\tilde{\mathbf{B}})$ is restrictive --- to be conjugate it needs to have the Kronecker product structure $\tilde{\vSigma}\otimes \tilde{\bV}$, which implies cross-equation restrictions on the covariance matrix. In particular, this structure requires symmetric treatment of own lags and lags of other variables. In many situations one might want to shrink the coefficients on lags of other variables more strongly to zero than those of own lags. This prior belief, however, cannot be implemented using the natural conjugate prior due to the Kronecker structure.

Here we develop a prior that solves this dilemma: this new prior allows asymmetric treatment between own lags and lags of other variables, while it maintains many useful analytical results. In what follows, we first consider a reparameterization of the reduced-form VAR in \eqref{eq:VAR}. We introduce in Section~\ref{ss:ACP} the new asymmetric conjugate prior and discuss its properties. We then derive the associated posterior distribution and discuss an efficient sampling scheme in Section~\ref{ss:post}. Finally, we give an analytical expression of the marginal likelihood in Section~\ref{ss:ML}.

\subsection{The Bayesian VAR in Structural Form} \label{ss:SVAR}

In this section we introduce a reparameterization of the reduced-form VAR in \eqref{eq:VAR} and derive the associated likelihood function. 
To that end, we first write the VAR in the following structural form:
\begin{equation} \label{eq:SVAR}
    \bA \by_t = \mathbf{b} + \bB_{1} \by_{t-1} + \cdots + \bB_{p} \by_{t-p} + \vepsilon_t^y, \quad \vepsilon_t^y \sim \distn{N}(\mathbf{0}, \vSigma),
\end{equation}
where $\mathbf{b}$ is an $n\times 1 $ vector of intercepts, $\bB_{1}, \ldots, \bB_{p}$ are $n \times n$ VAR coefficient matrices, $\bA$ is an $n \times n$ lower triangular matrix with ones on the diagonal and $\vSigma = \diag(\sigma_1^2, \ldots, \sigma_n^2)$ is diagonal. Since the covariance matrix $\vSigma$ is diagonal, we can estimate this recursive system equation by equation without loss of efficiency.\footnote{\citet*{CCM19} pioneer a similar equation-by-equation estimation approach to estimate a large VAR with a standard stochastic volatility specification. However, they use the reduced-form parameterization in \eqref{eq:VAR}, whereas here we use the structural form in \eqref{eq:SVAR}. As we will see below, the latter parameterization has the advantage of having a convenient representation as $n$ independent regressions and it consequently leads to a more efficient sampling scheme. \citet{AZ10} also consider a similar reparameterization of the reduced-form VAR that allows equation-by-equation estimation. But in their implementation they need to switch between two parameterizations, which makes estimation more cumbersome.} It is easy to see that we can recover the reduced-form parameters by setting $\tilde{\mathbf{b}} = \bA^{-1} \mathbf{b}$, $\tilde{\bB}_{j} =  \bA^{-1}\bB_{j}, j=1,\ldots, p$ and $\tilde{\vSigma} =  \bA^{-1} \vSigma (\bA^{-1})'$.

For later reference, we introduce some notations. Let $b_{i}$ denote the $i$-th element of $\mathbf{b}$ and let $\mathbf{b}_{j,i}$ represent the $i$-th row of $\bB_{j}$. Then,
$\vbeta_{i} = (b_{i},\mathbf{b}_{1,i},\ldots,\mathbf{b}_{p,i})'$ is the intercept and VAR coefficients for the $i$-th equation. Furthermore, let $\valpha_{i} $ denote the free elements
in the $i$-th row of the impact matrix $\bA$, i.e., $\valpha_{i} = (A_{i,1},\ldots, A_{i,i-1})'$. 
We then follow \citet{CE18b} to rewrite the $i$-th equation of the system in \eqref{eq:SVAR} as:
\[
    y_{i,t} = \tilde{\bw}_{i,t}\valpha_{i} + \tilde{\bx}_t \vbeta_{i}  + \epsilon_{i,t}^y, \quad \epsilon_{i,t}^y \sim \distn{N}(0, \sigma_i^2),
\]
where $\tilde{\bw}_{i,t} = (-y_{1,t},\ldots, -y_{i-1,t})$ and $\tilde{\bx}_t = (1, \by_{t-1}',\ldots, \by_{t-p}')$.
Note that $y_{i,t}$ depends on the contemporaneous variables $y_{1,t},\ldots, y_{i-1,t}$. But since the system is triangular, when we perform the change of variables from
$\vepsilon_t^y$ to $\by_t$  to obtain the likelihood function, the corresponding Jacobian has unit determinant and the likelihood function has the usual Gaussian form.

If we let $\bx_{i,t} = (\tilde{\bw}_{i,t}, \tilde{\bx}_t)$, we can further simplify the $i$-th equation as:
\[
	y_{i,t} = \bx_{i,t} \vtheta_{i} + \epsilon_{i,t}^y, \quad \epsilon_{i,t}^y \sim \distn{N}(0, \sigma_i^2),
\]
where $\vtheta_{i} = (\vbeta_{i}',\valpha_{i}')'$ is of dimension $k_i = np+i.$ Hence, we have rewritten the structural VAR in \eqref{eq:SVAR} as a system of $n$ independent regressions. Moreover, by stacking the elements of the impact matrix $\valpha_{i}$ and  the VAR coefficients $\vbeta_{i}$, we can sample them together to improve efficiency.\footnote{This more efficient blocking scheme has been used previously in the literature. For example, \citet*{ECS16} use it to speed up computations in the context of time-varying parameter VARs with stochastic volatility.} 

To derive the likelihood function, we further stack $\by_i = (y_{i,1},\ldots, y_{i,T})'$ and define $\bX_i$ and $ \vepsilon_{i}^y$ similarly. Hence, we can rewrite the above equation as follows:
\[
	\by_{i} = \bX_i \vtheta_{i} + \vepsilon_{i}^y, \quad \vepsilon_{i}^y \sim \distn{N}(\mathbf{0}, \sigma_i^2\mathbf{I}_T).
\]
Finally, let $\vtheta = (\vtheta_1',\ldots,\vtheta_n')'$ and $\vsigma^2=(\sigma_1^2,\ldots, \sigma_n^2)'$. Then, the likelihood function of the VAR in \eqref{eq:SVAR} is given by
\begin{equation}\label{eq:y_den}
	p(\by\gvn\vtheta, \vsigma^2) = \prod_{i=1}^n p(\by_i\gvn\vtheta_i, \sigma_i^2) = \prod_{i=1}^n  (2\pi\sigma_i^2)^{-\frac{T}{2}}\e^{-\frac{1}{2\sigma_i^2}(\by_i-\bX_i\vtheta_i)'(\by_i-\bX_i\vtheta_i)}.
\end{equation}
In other words, the likelihood function is the product of $n$ Gaussian densities. 

\subsection{Asymmetric Conjugate Priors} \label{ss:ACP}

Next we introduce a conjugate prior on $(\vtheta,\vsigma^2)$ that allows differential treatment between prior variances on own lags versus others. 
We assume that the parameters are {\em a priori} independent across equations, i.e., $p(\vtheta,\vsigma^2) = \prod_{i=1}^n p(\vtheta_i,\sigma^2_i)$. Furthermore, we consider a normal-inverse-gamma prior for each pair $(\vtheta_i,\sigma^2_i), i=1,\ldots, n$: 
\begin{equation}\label{eq:ACP}
	(\vtheta_i\gvn\sigma^2_i) \sim \distn{N}(\bm_i,\sigma^2_i\bV_{i}), \qquad \sigma^2_i \sim \distn{IG}(\nu_{i}, S_{i}),
\end{equation}
and we write $(\vtheta_i,\sigma^2_i)\sim\distn{NIG}(\bm_{i},\bV_{i},\nu_{i}, S_{i})$. In other words, the prior density of $(\vtheta,\vsigma^2)$ is given by
\begin{equation}\label{eq:prior_den}
	p(\vtheta,\vsigma^2) =\prod_{i=1}^n c_i	(\sigma^2_i)^{-\left(\nu_i+1+\frac{k_i}{2}\right)}
			\e^{-\frac{1}{\sigma^2_i}\left(S_i + \frac{1}{2}(\vtheta_i-\bm_i)'\bV_i^{-1}(\vtheta_i-\bm_{i})\right)},
\end{equation}
where $c_i = (2\pi)^{-\frac{k_i}{2}}|\bV_i|^{-\frac{1}{2}}S_i^{\nu_i}/\Gamma(\nu_i)$.\footnote{The proposed prior is related to the work of \citet{BH15}, who also consider equation-specific conjugate priors on the structural VAR coefficients and error variances. But since they consider a more general setting with a possibly non-triangular impact matrix $\bA$, they need a Metropolis-Hastings step to explore the marginal posterior distribution $\bA$. In contrast, in our special case with a lower triangular $\bA$, the proposed prior can be shown to be conjugate and direct sampling of all the parameters is available.} 

Since the prior variance of each element of $\vtheta_i$ is controlled by the corresponding diagonal element of $\bV_{i}$, it is obvious that this prior can accommodate different prior variances between own lags versus others. As we will show in the next section, this prior is also conjugate. To distinguish this from the natural conjugate prior, we call the prior in \eqref{eq:ACP} the asymmetric conjugate prior. The hyperparameters of the asymmetric conjugate prior are $\bm_i,\bV_i,\nu_i$ and $S_i$, $i=1,\ldots, n$. Next we describe how one can elicit these hyperparameters.

First partition $\bm_i = (\bm_{\vbeta,i}',\bm_{\valpha,i}')$ and $\bV_i = \text{diag}(\bV_{\vbeta,i},\bV_{\valpha,i})$, where $\bm_{\valpha,i}$ and  $\bV_{\valpha,i}$ are the hyperparameters
corresponding to $\valpha_i$, whereas  $\bm_{\vbeta,i}$ and  $\bV_{\vbeta,i}$ are those associated with $\vbeta_i$. In what follows, we first discuss eliciting the hyperparameters associated with 
$\valpha_i$ and $\sigma^2_i$ --- i.e.,  $\bm_{\valpha,i}$,  $\bV_{\valpha,i}, \nu_i$ and $S_i$. We then introduce two Minnesota-type shrinkage priors for the VAR coefficients $\vbeta_i$.


Since $\valpha_i$ and $\sigma^2_i$ control the reduced-form error covariance matrix $\tilde{\vSigma} = \bA^{-1} \vSigma (\bA^{-1})'$, one concern is that an arbitrary choice of the hyperparameters for $\valpha_i$ and $\sigma^2_i$ would induce some unreasonable prior on $\tilde{\vSigma}$. For example, the implied prior on the $i$-th diagonal element of $\tilde{\vSigma}$ might mechanically depend on its position in the $n$-tuple  --- i.e., the induced prior on 
$\tilde{\vSigma}$ depends on the order of the variables and is not invariant to reordering. This problem is especially acute for large systems due to the fact that $\bA^{-1}$ is lower triangular.

To avoid this potential non-invariance problem, we instead specify a prior on the reduced-form error covariance matrix $\tilde{\vSigma}$. And given this prior on  $\tilde{\vSigma}$, we 
then derive the implied prior on $\valpha_i$ and $\sigma^2_i, i=1,\ldots, n$. Since we can directly elicit prior beliefs on the elements of $\tilde{\vSigma}$, as a result these prior beliefs do not depend on the parameters' position in $\tilde{\vSigma}$. To that end, we consider a standard inverse-Wishart prior on $\tilde{\vSigma}$ centered around $\bS = \text{diag}(s_1^2,\ldots, s_n^2)$, where $s_i^2$ denotes the sample variance of the residuals from an AR(4) model for the variable~$i, i=1,\ldots, n$. More precisely, $\tilde{\vSigma} \sim \distn{IW}(\nu_0, \bS)$ with $\nu_0 = n + 2$. This prior on $\tilde{\vSigma}$ is commonly used in the literature \citep*[e.g., in][]{KK97,CCM15}. It turns out that, quite remarkably, the implied prior on $\valpha_i$ and $\sigma^2_i$ is normal-inverse-gamma. The following proposition and corollary summarize this result.\footnote{In the context of a general structural VAR, \citet{SZ98} suggest precisely this approach of deriving the implied prior on the impact matrix from a natural prior (inverse-Wishart) on the error covariance matrix. However, under their parameterization (Cholesky factorization), the implied prior on the impact matrix is non-standard. In contrast, we use a modified Cholesky factorization and the implied prior can be shown to be of the form of a normal-inverse-gamma distribution.}

\begin{proposition} \label{thm:main}  Consider the following normal-inverse-gamma priors on the diagonal elements of $\vSigma$ and the lower triangular elements of $\bA$:
\begin{align}
	\sigma_i^{2} & \sim \distn{IG}\left(\frac{\nu_0+i-n}{2},\frac{s_i^2}{2}\right), \quad i=1,\ldots, n, \label{eq:prior_sig2} \\	
	(A_{i,j}\gvn \sigma_i^{2}) & \sim \distn{N}\left(0, \frac{\sigma_i^2}{s_j^2}\right), \quad 1\leq j<i\leq n, \; i=2,\ldots, n. \label{eq:prior_Aij}
\end{align}
Then, $\tilde{\vSigma}^{-1}  = \bA' \vSigma^{-1} \bA$ has the Wishart distribution $\tilde{\vSigma}^{-1} \sim \mathcal{W} (\nu_0, \bS^{-1})$, where $ \bS = \text{diag}(s_1^2,\ldots, s_n^2)$.
It follows that $\tilde{\vSigma} \sim \distn{IW}(\nu_0, \bS)$. 
\end{proposition}

The proof is given in Appendix~C. Since the mapping  $\tilde{\vSigma}^{-1}  = \bA' \vSigma^{-1} \bA$ is one-to-one, the converse of Proposition~\ref{thm:main} is also true.

\begin{corollary} \label{thm:cor1}  Using the same notations as in Proposition~\ref{thm:main}, if $\tilde{\vSigma} \sim \distn{IW}(\nu_0, \bS)$, then the implied priors on $A_{i,j}$ 
and $\sigma^2_i$, $i=1,\ldots, n, j=1,\ldots, i-1$, are the normal-inverse-gamma distributions given in \eqref{eq:prior_sig2} and \eqref{eq:prior_Aij}.
\end{corollary}

The proof is given in Appendix~C. With these results, we can now make precise the claim that the proposed normal-inverse-gamma prior is independent of the order of the variables. Suppose the covariance matrix of the reduced-form error vector $\tilde{\vepsilon}_t^y$ is $\tilde{\vSigma}$, which has the prior $\distn{IW}(\nu_0,\bS)$. We claim that if we change the order of the reduced-form errors, we can use the normal-inverse-gamma prior to specify an appropriate inverse-Wishart prior --- we simply need to reorder the associated hyperparameters accordingly. More specifically, let $\pi$ denote an arbitrary permutation of $n$ elements with the associated permutation matrix $\bP_{\pi}$. If we permute the order of the dependent variables via 
$\bP_{\pi}\by_t = (y_{\pi(1),t},\ldots,y_{\pi(n),t})'$, its error covariance matrix is then $\bP_{\pi}\tilde{\vSigma}\bP_{\pi}'$ and has the $\distn{IW}(\nu_0, \bS_{\pi})$ distribution, where $\bS_{\pi} = \bP_{\pi}\bS\bP'_{\pi} = \text{diag}(s^2_{\pi(1)},\ldots,s^2_{\pi(n)}).$ Using Proposition~\ref{thm:main}, we can find a set of normal-inverse-gamma priors that induces this inverse-Wishart prior $\distn{IW}(\nu_0, \bS_{\pi}).$ We summarize this result in the following corollary. 

\begin{corollary} \label{thm:cor2} Suppose $\tilde{\vSigma} \sim \distn{IW}(\nu_0, \bS)$ and let
$\pi$ denote a permutation of $n$ elements with the associated permutation matrix $\bP_{\pi}$. It follows that $\bP_{\pi}\tilde{\vSigma}\bP_{\pi}' \sim \distn{IW}(\nu_0, \bS_{\pi}),$ where $\bS_{\pi} =\text{diag}(s^2_{\pi(1)},\ldots,s^2_{\pi(n)}).$ Consider the following normal-inverse-gamma priors on the diagonal elements of $\vSigma$ and the lower triangular elements of $\bA$:
\begin{align*}
	\sigma_i^{2} & \sim \distn{IG}\left(\frac{\nu_0+i-n}{2},\frac{s_{\pi(i)}^2}{2}\right),
	\quad i=1,\ldots, n,  \\	
	(A_{i,j}\gvn \sigma_i^{2}) & \sim \distn{N}\left(0, \frac{\sigma_i^2}
	{s_{\pi(j)}^2}\right), \quad 1\leq j<i\leq n, \; i=2,\ldots, n.
\end{align*}
Then, $\breve{\vSigma}^{-1}  = \bA' \vSigma^{-1} \bA$ has the Wishart distribution $\mathcal{W} (\nu_0, \bS_{\pi}^{-1})$. That is, $\breve{\vSigma} \sim \distn{IW}(\nu_0, \bS_{\pi})$. 

\end{corollary}

The proof follows directly from Proposition~\ref{thm:main}. In addition, Proposition~\ref{thm:main} holds for the more general case where $\bS$ is any symmetric positive definite matrix. That is, the induced priors on the structural-form variances are independent gamma distributions and the conditional priors of the free elements of $\bA$ are normal distributions. But unlike the previous case with diagonal $\bS$, here the free elements in the same row of $\bA$ are correlated. We summarize the results in the following corollary.\footnote{\citet{CJ09b} have shown a similar result. However, their proof is not sufficiently constructive and they did not give an explicit mapping between the inverse-Wishart parameters and the normal-inverse-gamma parameters.} Its proof is given in Appendix~C.

\begin{corollary} \label{thm:cor3} Suppose $\tilde{\vSigma} \sim \distn{IW}(\nu_0, \bR)$, where $\bR$ is a symmetric positive definite matrix. Factor 
$\tilde{\vSigma}^{-1}  = \bC' \vSigma^{-1} \bC$ and $\bR^{-1}  = \bL' \bS^{-1} \bL$, where $\bC$ and $\bL$ are lower triangular matrices with ones on the main diagonal, 
$ \vSigma=\text{diag}(\sigma_1^2,\ldots,\sigma_n^2)$ and $\bS =\text{diag}(s_1^2,\ldots,s_n^2) $ are diagonal matrices. Let  $\bc_{i}$ denote the free elements of the $i$-th row of $\bC$, i.e.,
$\bc_{i} = (C_{i,1},\ldots, C_{i,i-1})'$, and let $ \bL_{1:i-1}$ denote the $(i-1)\times(i-1)$ matrix that consists of the first $(i-1)$ rows and columns of $\bL$. Similarly define $\mathbf{l}_i$ and
$\bS_{1:i-1}$. Then, the implied priors on $\bc_{i}$ and $\sigma^2_i$ are 
\begin{align}
	\sigma_i^{2} & \sim \distn{IG}\left(\frac{\nu_0+i-n}{2},\frac{s_i^2}{2}\right), \quad i=1,\ldots, n, \label{eq:prior2_sig2} \\	
	(\bc_{i}\gvn \sigma_i^{2}) & \sim \distn{N}\left(\mathbf{l}_{i}, \sigma_i^2 \bL_{1:i-1}'\bS_{1:i-1}^{-1}\bL_{1:i-1} \right), 	\quad i=2,\ldots, n. \label{eq:prior2_Cij}
\end{align}
Since the mapping $\tilde{\vSigma}^{-1}  = \bC' \vSigma^{-1} \bC$  is one-to-one, the converse is also true. That is, if we assume that $(\bc_{i}, \sigma_i^{2})$ follows the normal-inverse-gamma distributions given in \eqref{eq:prior2_sig2} and \eqref{eq:prior2_Cij}, then the implied prior on $\tilde{\vSigma}$ is inverse-Wishart: $\tilde{\vSigma} \sim \distn{IW}(\nu_0, \bR)$.

\end{corollary}

Hence, the above proposition and corollaries give us a guide to elicit the hyperparameters associated with $\valpha_i$ and $\sigma^2_i$. In our baseline case, we set $\nu_i = 1 + i/2$, $S_i = s_i^2/2$, $\bm_{\valpha,i} = \mathbf{0}$ and $\bV_{\valpha,i} = \text{diag}(1/s_1^2,\ldots, 1/s_{i-1}^2)$. These values imply that the prior on the reduced-form error covariance matrix is $\tilde{\vSigma} \sim \distn{IW}(n + 2, \bS)$ with prior mean $\bS$. Furthermore, if we wish to have an inverse-Wishart prior on $\tilde{\vSigma} $ with a non-diagonal prior mean, we can simply 
use Corollary \ref{thm:cor3} to elicit the associated values for $\bm_{\valpha,i}$ and $\bV_{\valpha,i}$.

Next, we describe two ways to set the hyperparameters $\bm_{\vbeta,i}$ and  $\bV_{\vbeta,i}$. The first approach directly elicits prior beliefs on the structural VAR coefficients $\vbeta_i$. The idea of cross-variable shrinkage in this setting has been previously explored in \citet*{LSZ96} and \citet{SZ98}. We therefore follow \citet{SZ98}, who consider Minnesota-type shrinkage priors for VAR coefficients in the structural form.\footnote{As noted by \citet{SZ98}, a general structural VAR is a 
simultaneous equations model and there is no dependent variable in an equation (other than an arbitrary normalization). Hence, the distinction between `own' lags vs `others' is not meaningful. 
However, our setup is the special case of a recursive system, where there is a natural dependent variable in each equation. Hence, we can distinguish between `own' vs `other' lags.} More specifically, we set  $\bm_{\vbeta,i} = \mathbf{0}$ to shrink the VAR coefficients to zero for growth rates data; for level data,  $\bm_{\vbeta,i}$ is set to be zero as well except the coefficient associated with the first own lag, which is set to be one. 

Recall that  $\bV_{\vbeta,i}$ is the ratio of the prior covariance matrix of $\vbeta_i$ relative to the error variance $\sigma^2_i$. Similar to the Minnesota prior, here we assume  $\bV_{\vbeta,i}$ to be diagonal with the $k$-th diagonal element $(\bV_{\vbeta,i})_k$  set to be:
\[
	(\bV_{\vbeta,i})_k = \left\{
	\begin{array}{ll}
			\frac{\kappa_1}{l^2 s_i^2}, & \text{for the coefficient on the $l$-th lag of variable } i,\\
			\frac{\kappa_2}{l^2 s_j^2}, & \text{for the coefficient on the $l$-th lag of variable } j, j\neq i, \\			\
			\kappa_3, & \text{for the intercept}. \\
	\end{array} \right.	
\]
The hyperparameter $\kappa_1$ controls the overall shrinkage strength for coefficients on own lags, whereas $\kappa_2$ controls those on lags of other variables. These two hyperparameters will play a key role in the empirical analysis, and we will select them optimally by maximizing the associated marginal likelihood. We set $\kappa_3 = 100$, which implies essentially no shrinkage for the intercepts.\footnote{In principle one can select $\kappa_3 $ optimally as well, but the corresponding optimization is more costly to solve. More generally, high-dimensional numerical optimization using derivative-free methods is time consuming. One feasible alternative is to use Automatic Differentiation to obtain the relevant partial derivatives, which are then fed to numerical optimization routines that use these partial derivatives to more efficiently find the maximizer. See \citet{CJZ20} for an example.}

By contrast, in the second approach we first elicit the prior means and variances on the reduced-form VAR coefficients. We then derive the implied prior means and variances on the structural-form VAR coefficients. Since the prior beliefs are elicited on the reduced-form parameters, this approach is comparable to commonly-used Minnesota priors on the reduced-form parameters. More specifically, let $\tilde{\bm}_{\vbeta,i}$ and  $\sigma_{i}^2\tilde{\bV}_{\vbeta,i}$ denote the prior mean vector and covariance matrix of the reduced-form parameters $\tilde{\vbeta}_i$ --- these hyperparameters can be elicited similarly as above. For later reference, we let $\tilde{\kappa}_1$ and $\tilde{\kappa}_2$ denote the corresponding shrinkage hyperparameters. Finally, let $\bm_{\vbeta,i}$ and  $\sigma_{i}^2\bV_{\vbeta,i}$ denote the corresponding hyperparameters of the structural-form parameters $\vbeta_i$. Appendix D provides the details of the derivation and the explicit formulas of $\bm_{\vbeta,i}$ and  $\bV_{\vbeta,i}$ using $\tilde{\bm}_{\vbeta,i}$ and  $\tilde{\bV}_{\vbeta,i}$ as inputs. Both approaches give a normal-inverse-gamma prior of the form given in~\eqref{eq:ACP}.

\subsection{Posterior Distribution and Efficient Sampling} \label{ss:post}

In this section we first derive the posterior distribution of $(\vtheta,\vsigma^2)$ under the asymmetric conjugate prior and show that it has indeed the same form as the prior. Then, we describe an efficient method for posterior simulation. 

Since both the likelihood in \eqref{eq:y_den} and the prior in \eqref{eq:prior_den} have the product form, 
we can estimate each pair $(\vtheta_i,\sigma_i^2)$ separately. More specifically, the posterior distribution of $(\vtheta,\vsigma^2)$ is given by:
\begin{align*}
	p(\vtheta,& \vsigma^2\gvn \by) \propto  p(\vtheta,\vsigma^2)p(\by\gvn\vtheta,\vsigma^2)  = \prod_{i=1}^n  p(\vtheta_i,\sigma^2_i)p(\by_i\gvn\vtheta_i,\sigma_i^2)\\
		& =  \prod_{i=1}^nc_i(\sigma^2_i)^{-\left(\nu_i+1+\frac{k_i}{2}\right)}
			\e^{-\frac{1}{\sigma^2_i}\left(S_i + \frac{1}{2}(\vtheta_i-\bm_i)'\bV_i^{-1}(\vtheta_i-\bm_i)\right)} \times
				(2\pi\sigma^2_i)^{-\frac{T}{2}}\e^{-\frac{1}{2\sigma^2_i}(\by_i-\bX_i\vtheta_i)'(\by_i-\bX_i\vtheta_i)} \\
		& = \prod_{i=1}^nc_i(2\pi)^{-\frac{T}{2}} (\sigma^2_i)^{-\left(\nu_i+\frac{T+k_i}{2}+1\right)}	
		\e^{-\frac{1}{\sigma^2_i}\left(S_i + \frac{1}{2}\left(\vtheta_i'(\bV_i^{-1} + \bX_i'\bX_i)\vtheta_i - 2\vtheta_i'(\bV_i^{-1}\bm_i+ \bX_i'\by_i)
		+ \bm_i'\bV_i^{-1}\bm_i	+  \by_i'\by_i\right)\right)} \\
		& = \prod_{i=1}^n c_i(2\pi)^{-\frac{T}{2}} (\sigma^2_i)^{-\left(\nu_i+\frac{T+k_i}{2}+1\right)}
			\e^{-\frac{1}{\sigma^2_i}\left(\hat{S}_i + \frac{1}{2}(\vtheta_i-\hat{\vtheta}_i)'\bK_{\vtheta_i}(\vtheta_i-\hat{\vtheta}_i)\right)}, 
\end{align*}
where $\bK_{\vtheta_i} = \bV_i^{-1} + \bX_i'\bX_i$, $\hat{\vtheta}_i = \bK_{\vtheta_i}^{-1}(\bV_i^{-1}\bm_i+ \bX_i'\by_i)$ and $\hat{S}_i = S_i + (\by_i'\by_i + \bm_i'\bV_i^{-1}\bm_i 
- \hat{\vtheta}_i'\bK_{\vtheta_i}\hat{\vtheta}_i)/2$. Hence, the posterior distribution is a product of $n$  normal-inverse-gamma distributions and we have:
\begin{equation}\label{eq:post_prod}
	(\vtheta_i,\sigma^2_i\gvn\by) \sim \distn{NIG}\left(\hat{\vtheta}_i,\bK_{\vtheta_i}^{-1},\nu_i + \frac{T}{2},\hat{S}_i\right), \quad i=1,\ldots, n.
\end{equation}

Using properties of the normal-inverse-gamma distribution, it is easy to see that the posterior means of $\vtheta_i$ and $\sigma^2_i$ are respectively $\hat{\vtheta}_i$ and 
$\hat{S}_i/(\nu_i + T/2 - 1)$. Other posterior moments can also be obtained by using similar properties of the normal-inverse-gamma distribution. 
For other quantities of interest where analytical results are not available, we can estimate them by posterior simulation. For example, the $h$-step-ahead predictive distribution of $\by_{T+h}$ is non-standard. But we can obtain posterior draws from $p(\vtheta, \vsigma^2\gvn \by)$ to construct the $h$-step-ahead predictive distribution. 

In what follows, we outline an efficient method to simulate a sample of size $M$ from the posterior distribution. Here we can directly generate {\em independent} draws from the posterior distribution as opposed to MCMC draws that are correlated by construction. First, note that $(\vtheta, \vsigma^2\gvn \by)$ is a product of $n$ normal-inverse-gamma distributions as given in~\eqref{eq:post_prod}. Thus we can sample each pair $(\vtheta_i,\sigma^2_i\gvn\by)$ individually. Next, we can sample $(\vtheta_i,\sigma^2_i\gvn\by)$ in two steps. First, we draw $\sigma^2_i$ marginally from $(\sigma^2_i\gvn \by) \sim\distn{IG}(\nu_i+T/2, \hat{S}_i)$. Then, given the $\sigma^2_i$ sampled, we obtain $\vtheta_i$ from the conditional distribution 
\[
	(\vtheta_i \gvn \by,\sigma_i^2) \sim\distn{N}(\hat{\vtheta}_i, \sigma^2_i \bK_{\vtheta_i}^{-1}).
\]
Here the covariance matrix $ \sigma^2_i \bK_{\vtheta_i}^{-1}$ is of dimension $k_i = np+i$. When $n$ is large, sampling from this normal distribution using conventional 
methods --- based on the Cholesky factor of $ \sigma^2_i \bK_{\vtheta_i}^{-1}$ --- is computationally intensive for two reasons. First, inverting the $k_i\times k_i$ matrix $\bK_{\vtheta_i}$ 
to obtain the covariance matrix $\sigma^2_i \bK_{\vtheta_i}^{-1}$ is computationally costly. Second, the Cholesky factor of the covariance matrix needs to be computed $M$ times --- once for each draw of $\sigma^2_i$ from the marginal distribution. It turns out that both of these computationally intensive steps can be avoided.

To that end, we introduce the following notations: given a non-singular square matrix $\bF$ and a conformable vector $\mathbf{d}$, let $\bF\backslash \mathbf{d}$ denote the unique solution to 
the linear system $\bF \bz = \mathbf{d}$, i.e., $\bF\backslash \mathbf{d} = \bF^{-1}\mathbf{d}$. When $\bF$ is lower triangular, this linear system can be solved quickly by forward substitution; when $\bF$ is upper triangular, it can be solved by backward substitution.\footnote{Forward and backward substitutions are implemented in standard packages such as \textsc{Matlab}, \textsc{Gauss} and \textsc{R}. In  \textsc{Matlab}, for example, it is done by \texttt{mldivide($\mathbf{F},\mathbf{d}$)} or simply \texttt{$\mathbf{F} \backslash \mathbf{d}$}.} Now, compute the Cholesky factor $\bC_{\bK_{\vtheta_i}}$ of $\bK_{\vtheta_i}$ such that $\bK_{\vtheta_i} = \bC_{\bK_{\vtheta_i}}\bC_{\bK_{\vtheta_i}}'$. 
Note that this needs to be done only once. Let $\bu$ be a $k_i\times 1 $ vector of independent sample from $\distn{N}(0,\sigma_i^2)$. Then, return
\[
	\hat{\vtheta}_i + \bC_{\bK_{\vtheta}}'\backslash \bu,
\]
which has the $\distn{N}(\hat{\vtheta}_i, \sigma^2_i \bK_{\vtheta_i}^{-1})$ distribution.\footnote{Note that $\hat{\vtheta}_i$ can be obtained similarly without explicitly computing the inverse of $\bK_{\vtheta_i}$.
Specifically, it is easy to see that  $\hat{\vtheta}_i$ can be calculated as: $\bC_{\bK_{\vtheta_i}}'\backslash (\bC_{\bK_{\vtheta_i}} \backslash (\bV_i^{-1}\bm_i+ \bX_i'\by_i))$ by forward then backward substitution. Also note that since $\bV_i^{-1}$ is diagonal, its inverse is straightforward to compute.} Finally, we can further speed up the computations by vectorizing all operations to obtain $M$ posterior draws instead of using for-loops.

This sampling scheme is more efficient than the method in \citet{CCM19}, who propose estimating the reduced-form parameters equation-by-equation. The main reason is that their method requires computing the Cholesky factor of every sampled reduced-form error covariance matrix (e.g., a total of $M$ times for $M$ draws as they use MCMC). In contrast, the proposed method needs to compute the Cholesky factor of the error covariance matrix only once, because here it does not depend on the sampled coefficients (i.e., the proposed sampler is not a Gibbs sampler). This difference becomes more important when $n$ becomes larger, as number of operations for computing the Cholesky factor of an $n\times n$ matrix is $\mathcal{O}(n^3)$.

To get a sense of how long it takes to obtain posterior draws using the proposed algorithm, we fit Bayesian VARs of different sizes, each with $p=4$ lags. The algorithm is implemented using \textsc{Matlab} on a desktop with an Intel Core i7-7700 @3.60 GHz processor and 64GB memory. The computation times (in seconds) to obtain 10,000 posterior draws of $(\vtheta, \vsigma^2) $ are reported in Table~\ref{tab:times}. As it is evident from the table, the proposed method is fast and scales well. It also compares favorably to the algorithm in \citet{CCM19}, especially when $n$ is large. For example, for a large BVAR with 	$n=100$	variables, the proposed method takes about half a minute to obtain 10,000 posterior draws. In comparison, using the algorithm in \citet{CCM19} takes about 43 minutes.\footnote{As recently pointed out in \citet{Bognanni21}, the original algorithm in \citet{CCM19} is not exact and can only be viewed as an approximation. The corrected algorithm described in \citet{CCCM21} is about 3-10 times slower than the original algorithm, depending on the size of the system. Hence, the advantage of our posterior sampler is even more substantial when the corrected algorithm is used.} 

\begin{table}[H]
\centering
\caption{The computation times (in seconds) to obtain 10,000 posterior draws of $(\vtheta, \vsigma^2) $ using the proposed method compared to the method in  \citet{CCM19}. All BVARs have
$p=4$ lags.}  \label{tab:times}
\begin{tabular}{lccc}
\hline\hline	
	&	$n=25$	&	$n=50$	&	$n=100$	\\ \hline
proposed method	&	1.3	&	6.8	&	28	\\
\rowcolor{lightgray}
CCM	&	58	&	238	&	 2,574	\\
\hline\hline
\end{tabular}
\end{table}

\subsection{The Marginal Likelihood} \label{ss:ML}

In this section we provide an analytical expression of the marginal likelihood. This closed-form expression is useful for a range of purposes, 
such as obtaining optimal hyperparamaters or designing efficient estimation algorithms for more flexible Bayesian VARs.

To prevent arithmetic underflow and overflow, we evaluate the marginal likelihood in log scale. 
Given the likelihood function in \eqref{eq:y_den} and the asymmetric conjugate prior in \eqref{eq:prior_den}, the associated log marginal likelihood of the VAR
has the following analytical expression:
\begin{equation} \label{eq:lml}
\begin{split}
	\log p(\by) = & -\frac{Tn}{2}\log(2\pi)  + \sum_{i=1}^n \left[ -\frac{1}{2}\left(\log|\bV_i| + \log|\bK_{\vtheta_i}|\right) \right. \\
	& \left. + \log \Gamma\left(\nu_i+\frac{T}{2}\right) + \nu_i\log S_i - \log\Gamma(\nu_i) -\left(\nu_i+\frac{T}{2}\right)\log\hat{S}_i\right].
\end{split}
\end{equation}
The details of the derivation are given in Appendix B. The above expression is straightforward to evaluate. We only note that to compute the log determinant $\log|\bK_{\vtheta_i}|$, 
it is numerically more stable to first compute its Cholesky factor $\bC_{\bK_{\vtheta_i}}$ and return $2\sum \log c_{ii}$, where $c_{ii}$ is the $i$-th diagonal element of the $\bC_{\bK_{\vtheta_i}}$.

\section{Extensions} \label{s:extensions}

In this section we briefly discuss how we can use the above analytical results and the efficient sampling scheme in more general settings. Suppose we augment our BVAR in~\eqref{eq:y_den} 
to the model $p(\by\gvn\vtheta,\vsigma^2,\vgamma)$ with the additional parameter vector $\vgamma$. Further, consider the prior $p(\vtheta,\vsigma^2,\vgamma) = p(\vtheta,\vsigma^2\gvn \vgamma)p(\vgamma)$, where $p(\vtheta,\vsigma^2\gvn\vgamma)$ is the asymmetric conjugate prior that could potentially depend on $\vgamma$ and the marginal prior $p(\vgamma)$ is left unspecified for now. 
Before we discuss some efficient posterior samplers, we first give two examples that fit into this framework. 

In our first example, we augment the BVAR by treating the hyperparameters $\kappa_1$ and $\kappa_2$ as parameters to be estimated. That is, $\vgamma = (\kappa_1,\kappa_2)'$. 
This extension is useful as it takes into account the parameter uncertainty of  $\kappa_1$ and $\kappa_2$ \citep*[see also][]{GLP15}. This extension is considered in the empirical application. In our second example, we extend the BVAR by adding an MA(1) component to each equation:
\begin{align*}
	y_{i,t} & = \bx_{i,t} \vtheta_{i} + \epsilon_{i,t}^y,  \\
	\epsilon_{i,t}^y & = u_{i,t} + \psi_i u_{i,t-1}, 
\end{align*}
where $u_{i,t} \sim \distn{N}(0, \sigma_i^2), t=1,\ldots, T, i=1,\ldots, n$. In this case, $\vgamma=(\psi_1,\ldots, \psi_n)'$. This extension is motivated by the empirical finding that allowing for moving average errors often improves
forecast performance \citep[see, e.g.,][]{chan13, chan20}.
 
Both examples fit into the framework with likelihood $p(\by\gvn\vtheta,\vsigma^2,\vgamma)$ and prior $p(\vtheta,\vsigma^2,\vgamma)$. 
One natural posterior sampler is to construct a Markov chain by sequentially sampling from $p(\vtheta,\vsigma^2\gvn \by,\vgamma)$ and $p(\vgamma \gvn \by, \vtheta,\vsigma^2)$. 
The first density is a product of normal-inverse-gamma densities, and we can efficiently obtain a draw from it as described before. 
The second density depends on the model, but it is often easy to sample from.\footnote{For our first example with a low-dimensional $\vgamma = (\kappa_1,\kappa_2)'$, 
an independent-chain Metropolis-Hastings algorithm can be easily constructed. For our  second example with $\vgamma=(\psi_1,\ldots, \psi_n)'$, it turns out that we can factor 
$p(\vgamma \gvn \by, \vtheta,\vsigma^2) = p(\vpsi \gvn \by, \vtheta,\vsigma^2) = \prod_{i=1}^n p(\psi_i \gvn \by, \vtheta,\vsigma^2)$. Then, each $\psi_i$ can be simulated using the 
method in \citet{chan13}.}

Alternatively, a more efficient approach is the collapsed sampler that samples from $p(\vgamma \gvn \by)$. This sampling scheme is typically more efficient as it integrates out the high-dimensional parameters $(\vtheta,\vsigma^2)$ analytically.
The density  $p(\vgamma \gvn \by)$ can be evaluated quickly since
\[
	p(\vgamma \gvn \by) \propto p(\by \gvn \vgamma) p(\vgamma),
\]
where $p(\by \gvn \vgamma) $ is the `marginal likelihood' of the standard BVAR. For instance, for the first example with $\vgamma = (\kappa_1,\kappa_2)'$, the 
quantity $p(\by \gvn \vgamma)$ is exactly as the analytical expression given in \eqref{eq:lml}. Finally, given the posterior draws of $\vgamma$, we can obtain the posterior draws 
of $(\vtheta,\vsigma^2)$ from $p(\vtheta,\vsigma^2\gvn \by,\vgamma)$.
 
\section{Application: Identifying Financial Shocks}  \label{s:application}

In this section we illustrate the usefulness of the proposed asymmetric conjugate prior by revisiting the empirical study in \citet{FRS19}, who identify financial shocks in a VAR using sign restrictions. More specifically, for their baseline they consider a 6-variable VAR --- using US data on GDP, prices, interest rate, investment, stock prices and credit spread --- to identify demand, supply, monetary, investment and financial shocks using a set of sign restrictions on the contemporaneous impact matrix. Here we revisit their empirical study by using a larger set of variables and consider a 15-variable VAR. 

There are a few reasons in favor of using a larger set of macroeconomic and financial variables. First, there is the concern of informational deficiency of using a limited information set. As pointed out in the seminal papers by \citet{HS91} and \citet{LR93, LR94}, when the econometrician considers a narrower set of variables than the economic agent, the underlying model used by the econometrician is non-fundamental, in the sense that current and past observations of the variables do not span the same space spanned by the structural shocks. Consequently, structural shocks cannot be recovered from the model. Using a larger set of relevant variables can alleviate this concern \citep[see, e.g.,][for a recent review on non-fundamentalness]{gambetti21}.

Second, the mapping from variables in an economic model to the data is typically not unique. For example, take the economic variable inflation. Should one match it to the data based on the CPI, PCE, or the GDP deflator \citep{LMW20}? One natural way to circumvent an arbitrary choice is to include multiple data series corresponding to the same economic variable in the analysis. An added benefit of this approach is that by using more data series (and the corresponding sign restrictions), one often obtains sharper inference. 

While \citet{FRS19} consider a non-informative prior for their 6-variable VAR, shrinkage is essential here for our much larger system. In particular, we use the proposed asymmetric conjugate prior to shrink the VAR coefficients in a data-based manner. After describing the macroeconomic dataset in Section~\ref{ss:data}, we first present the estimates of the shrinkage hyperparameters in Section~\ref{ss:fullsample}, highlighting the 
empirical relevance of allowing for different levels of shrinkage on own lags and other lags. We then conduct a structural analysis with sign restrictions in Section~\ref{ss:IR}.

\subsection{Data} \label{ss:data}

The dataset for our empirical application consists of 15 US quarterly variables and the sample period is from 1985:Q1 to 2019:Q4. These variables are constructed from raw time-series obtained from various sources, including the FRED database at the Federal Reserve Bank of St. Louis and the Federal Reserve Bank of Philadelphia. The complete list of these time-series and their sources are given in Appendix~A. 

In addition to the 6 variables used in the baseline model in \citet{FRS19} --- namely, real GDP, GDP deflator, 3-month treasury rate, ratio of private investment over output, S\&P 500 index and a credit spread defined as the difference between Moody's baa corporate bond yield and the federal funds rate--- we include 9 additional macroeconomic and financial variables, such as ratio of total credit over real estate value, industrial production, mortgage rates, as well as other measures of inflation, interest rates and stock prices. These 15 variables are listed in Table~\ref{tab:sign}.

\subsection{Optimal Shrinkage Hyperparameters} \label{ss:fullsample}

In this section we fit the 15-variable VAR with the proposed asymmetric conjugate prior. More specifically, we implement the version that elicits prior beliefs on the reduced-form parameters $\tilde{\vbeta}_i$ with hyperparameters $\tilde{\kappa}_1$ and $\tilde{\kappa}_2$. We obtain the optimal hyperparameter values by maximizing the log marginal likelihood. The results are reported in Table~\ref{tab:kappas}. 

For comparison, we also consider two useful benchmarks. In the first case we set $\tilde{\kappa}_1 = \tilde{\kappa}_2 = \tilde{\kappa}$ and maximize the log marginal likelihood with respect to $\tilde{\kappa}$ only. This benchmark mimics the standard practice of using the natural conjugate prior that does not distinguish between own lags and lags of other variables. We refer to this version as the symmetric prior. The second benchmark is a set of subjectively chosen values that apply cross-variable shrinkage. In particular, we follow \citet*{CCM15} and consider $\tilde{\kappa}_1 = 0.04$ and $\tilde{\kappa}_2 = 0.0016$. This second benchmark is referred to as the subjective prior. 

\begin{table}[H]
\centering
\caption{Optimal values of the hyperparameters under the symmetric prior $(\tilde{\kappa}_1=\tilde{\kappa}_2)$, the subjective prior \citep{CCM15} and the proposed asymmetric prior 
(elicited on the reduced-form parameters).} \label{tab:kappas}
\begin{tabular}{lccc}
\hline\hline	
	&	Symmetric prior	&	Subjective prior	&	Asymmetric prior \\ \hline
$\tilde{\kappa}_1$	&	0.008	&	0.040	&	0.058	\\
\rowcolor{lightgray}
$\tilde{\kappa}_2$	&	0.008	&	0.0016 &	0.0043	\\
log-ML	&	4,333.3	&	4,329.8 &	4,341.6	\\ \hline\hline
\end{tabular}
\end{table}

Under the symmetric prior with the restriction that $\tilde{\kappa}_1 = \tilde{\kappa}_2 $, the optimal hyperparameter value is 0.008. If we allow  $\tilde{\kappa}_1$ and $\tilde{\kappa}_2$ to be different, we obtain very different results: the optimal value for $\tilde{\kappa}_1$ increases more than 7 times to 0.058, whereas the optimal value of $\tilde{\kappa}_2$ reduces by about half to 0.0043. These results suggest that the data prefers shrinking the coefficients on lags of other variables much more aggressively to zero than those on own lags. This makes intuitive sense as one would expect, on average, a variable's own lags would contain more information about its future evolution than lags of other variables. By relaxing the restriction that $\tilde{\kappa}_1 = \tilde{\kappa}_2$, the marginal likelihood value increases by about 4,000. If we were to test the hypothesis that $\tilde{\kappa}_1 = \tilde{\kappa}_2$, this large difference in marginal likelihood values would have decidedly reject it.

In addition, the optimal values of $\tilde{\kappa}_1$ and $\tilde{\kappa}_2$ under the asymmetric prior are also quite different from those of the subjective prior with $\tilde{\kappa}_1 = 0.04$ and $\tilde{\kappa}_2 = 0.0016$. By selecting the values of $\tilde{\kappa}_1$ and $\tilde{\kappa}_2$ optimally, one can increase the marginal likelihood value by more than 13,000. These results suggest that the subjective prior shrinks both the coefficients on own and other lags too aggressively for our dataset.

\begin{figure}[H]
    \centering
   \includegraphics[width=1\textwidth]{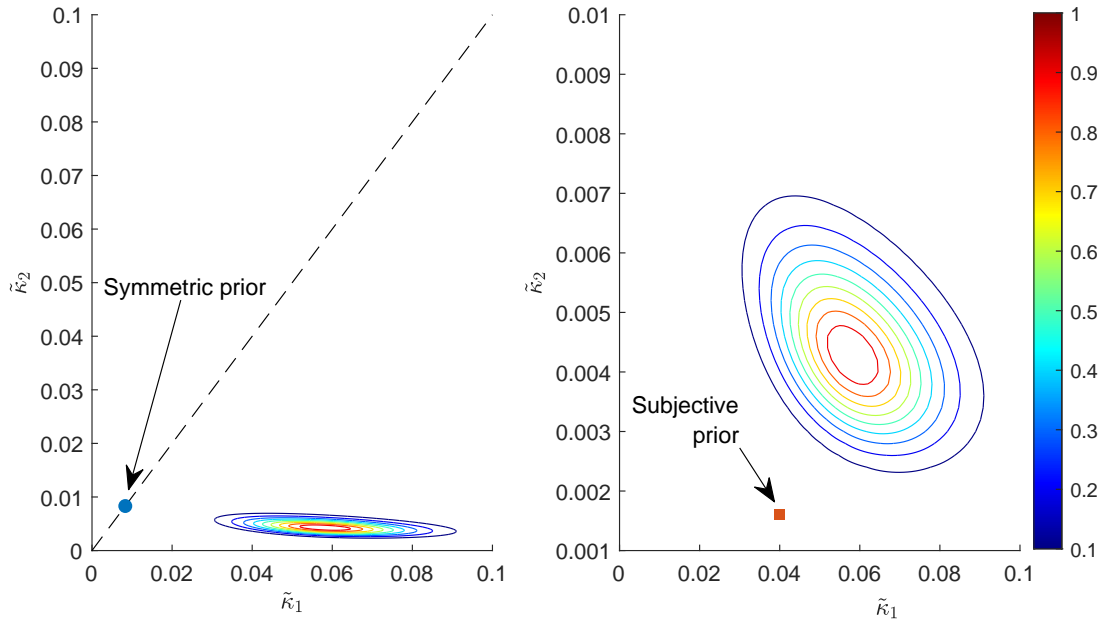}
   \caption{Contour plot of the joint posterior density of $\tilde{\kappa}_1$ and $\tilde{\kappa}_2$. The mode of the density is normalized to one for easy comparison.}
   \label{fig:post_kappa}	
\end{figure}

We have so far taken the empirical Bayes approach of choosing hyperparameter values by maximizing the log marginal likelihood. A fully Bayesian approach would specify proper priors on $\tilde{\kappa}_1$ and $\tilde{\kappa}_2 $ and obtain the corresponding posterior distribution. The latter approach has the additional advantage of being able to quantity parameter uncertainty of $\tilde{\kappa}_1$ and $\tilde{\kappa}_2$. 
In view of this, we take a fully Bayesian approach and treat $\tilde{\kappa}_1$ and $\tilde{\kappa}_2 $ as parameters to be estimated. Specifically, we assume a uniform prior on the unit square $(0,1)\times (0,1)$ for $\tilde{\kappa}_1$ and $\tilde{\kappa}_2$, and compute the marginal posterior distribution of $\tilde{\kappa}_1$ and $\tilde{\kappa}_2$. The contour plot of the joint posterior density is reported in Figure~\ref{fig:post_kappa}. 

As the contour plot in the right panel shows, most of the mass of $\tilde{\kappa}_1$ lies between 0.03 and 0.09, whereas the mass of $\tilde{\kappa}_2$ is mostly between 0.002 and 0.007. That is, $\tilde{\kappa}_1$ tends to be an order of magnitude larger than $\tilde{\kappa}_2$. These results confirm the conclusion that one should shrink the coefficients on other lags much more aggressively to zero than those on own lags. Moreover, since there is virtually no mass along the diagonal line $\tilde{\kappa}_1 = \tilde{\kappa}_2$, requiring them to be the same as in the natural conjugate prior appears to be too restrictive. 
As a comparison, we also plot the values of $\tilde{\kappa}_1$ and $\tilde{\kappa}_2$ under the symmetric and subjective priors. As is evident from the figure, the values under both priors are far from the high-density region of the posterior distribution.

Overall, the estimation results indicate that the optimal hyperparameter values could be very different from some subjectively chosen values commonly used in empirical work. In addition, they also highlight the importance of allowing for different levels of shrinkage on own versus other lags --- and therefore the empirical relevance of the proposed asymmetric conjugate prior.

\subsection{Sign Restrictions and Impulse Responses} \label{ss:IR}

Next, we revisit the empirical application in \citet{FRS19} that identifies financial shocks using sign restrictions on the contemporaneous impact matrix. We first replicate their baseline results from a 6-variable VAR to identify 5 structural shocks, namely, demand, supply, monetary, investment and financial shocks. We then consider a larger, 15-variable VAR with the asymmetric conjugate prior to identify the same structural shocks. 

For the replication exercise, we use the same 6 variables and the associated sign restrictions considered in \citet{FRS19}, which are listed in the first 6 rows of Table~\ref{tab:sign}. The sign restrictions to identify supply, demand and monetary shocks are standard and are consistent with a wide range of dynamic stochastic general equilibrium models \citep[see, e.g.,][]{CP11}. To disentangle investment and financial shocks from demand shocks, they are assumed to have different effects on the ratio of investment over output. More specifically, positive investment and financial shocks have a positive effect on the ratio, which is in line with the idea that these shocks create investment booms. In contrast, positive demand shocks reduce the ratio of investment over output --- investment level can still increase in response to demand shocks, but not as much as other components of the output. 

Finally, stock prices are used to disentangle investment shocks from financial shocks, following the influential paper by \citet{CMR14}. In particular, investment shocks are assumed to have a negative impact on the stock prices, whereas financial shocks have a positive impact. The first assumption captures the idea that investment shocks are shocks to the supply of capital, which generate negative comovements between the stock of capital and its price, where the price of capital is viewed as a proxy of the value of equity. In contrast, financial shocks are stocks to the demand for capital, which generate positive comovements between the stock of capital and its price. 

We then augment the 6-variable VAR with 9 additional variables. Some of the new variables are alternative time series corresponding to a particular economic variable (e.g., CPI and PCE as prices; the Dow Jones Industrial Average as stock prices). In those cases the same sign restrictions for the economic variable as discussed above are imposed. In addition, we also include a few other seemingly relevant variables to alleviate the concern of informational deficiency. The additional variables and the corresponding sign restrictions are listed in rows 7-15 of Table~\ref{tab:sign}.

Given the posterior draws of the parameters, the algorithm proposed in \citet{RWZ10} is used to incorporate the sign restrictions to construct impulse responses. This procedure can be computational intensive when the number of variables and the number of identified shocks are large. In particular, when there are a large number of sign restrictions, it is not uncommon to require millions of posterior draws to obtain enough samples that satisfy all the sign restrictions.\footnote{There are different implementations of the algorithm in \citet{RWZ10}. One common approach, as is done in 
\citet{FRS19}, is to combine every uniform draw from the orthogonal group $\text{O}(n)$ with a new posterior draw to generate candidate impulse responses. If the impulse responses do not satisfy all the sign restrictions, a new orthogonal matrix and a new posterior draw are used to generate candidate impulse responses. Alternatively, one can reuse the posterior draw and only generate a new orthogonal matrix. In this case the number of posterior draws required can be reduced. But to ensure the algorithm terminates, one typically sets an upper bound on the number of times a posterior draw can be reused.} For instance, \citet{FRS19} consider a non-informative prior and construct impulse responses using a standard Gibbs sampler in conjunction with the algorithm in \citet{RWZ10}. They report estimation time of about a week for the 6-variable VAR using a 12-core workstation. For our 15-variable VAR with more sign restrictions, it would be extremely computationally expensive to use a standard Gibbs sampler to obtain enough posterior draws. To make estimation feasible, we adopt the proposed asymmetric conjugate prior and the associated efficient sampler to generate posterior draws of the structural parameters. These structural parameters are then transformed to the corresponding reduced-form parameters, which are then used to construct candidate impulse responses.

\begin{table}[H]
\centering
\caption{Sign restrictions and identified shocks.} \label{tab:sign}
\begin{tabular}{lccccc}
\hline\hline
	&	Supply	&	Demand	&	Monetary	&	Investment	&	Financial	\\ \hline
GDP	&	$+$	&	$+$	&	$+$	&	$+$	&	$+$	\\
\rowcolor{lightgray}
GDP deflator	&	$-$	&	$+$	&	$+$	&	$+$	&	$+$	\\
3-month tbill rate	&	NA	&	$+$	&	$-$	&	$+$	&	$+$	\\
\rowcolor{lightgray}
Investment/output	&	NA	&	$-$	&	NA	&	$+$	&	$+$	\\
S\&P 500	&	$+$	&	NA	&	NA	&	$-$	&	$+$	\\
\rowcolor{lightgray}
Spread	&	NA	&	NA	&	NA	&	NA	&	NA	\\
Spread 2	&	NA	&	NA	&	NA	&	NA	&	NA	\\
\rowcolor{lightgray}
Credit/Real estate value	&	NA	&	NA	&	NA	&	NA	&	NA	\\
Mortgage rates	&	NA	&	NA	&	NA	&	NA	&	NA	\\
\rowcolor{lightgray}
CPI	&	$-$	&	$+$	&	$+$	&	$+$	&	$+$	\\
PCE	&	$-$	&	$+$	&	$+$	&	$+$	&	$+$	\\
\rowcolor{lightgray}
employment	&	NA	&	NA	&	NA	&	NA	&	NA	\\
Industrial production	&	$+$	&	$+$	&	$+$	&	$+$	&	$+$	\\
\rowcolor{lightgray}
1-year tbill rate	&	NA	&	$+$	&	$-$	&	$+$	&	$+$	\\
DJIA	&	$+$	&	NA	&	NA	&	$-$	&	$+$	\\
\hline\hline
\end{tabular}
{\raggedright \footnotesize{Note: Spread is defined as the difference between Moody's baa corporate bond yield and the federal funds rate; Spread 2 is the difference between Moody's baa corporate bond yield and 10-year treasury yield.} \par}
\end{table}

\begin{figure}[H]
    \centering
   \includegraphics[width=1\textwidth]{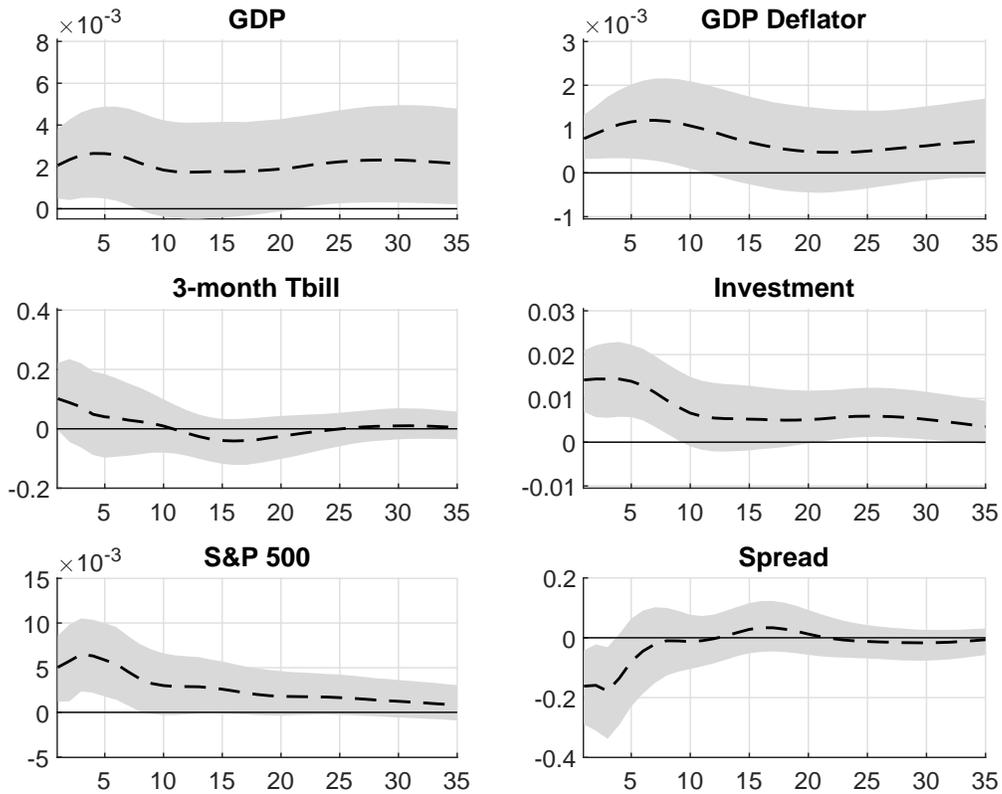}
   \caption{Impulse responses from a 6-variable VAR with the asymmetric conjugate prior to an one-standard-deviation financial shock. The shaded region represents the 16-th and 84-th percentiles.}
   \label{fig:IRF_finc_6}	
\end{figure}

We first report results from the baseline model in \citet{FRS19} using the asymmetric conjugate prior elicited on the reduced-form parameters and an updated dataset. Since an improper (non-informative) prior is used in the original study, to make our results comparable, we consider a proper but relatively vague prior by setting $\tilde{\kappa}_1 =\tilde{\kappa}_2 = 1$. Figure~\ref{fig:IRF_finc_6} plots the impulse responses of the 6 variables to an one-standard-deviation financial shock. 

Despite using a different prior and dataset, our results are remarkably similar to those reported in Figure 1 of \citet{FRS19}. In particular, we also find a large effect on GDP and a relatively smaller impact on prices. The responses of investment and stock prices are persistent. In particular, the credible intervals exclude 0 for the first~7 quarters after impact for both variables. It is also interesting to note that even though the responses of the credit spread are unrestricted in the estimation, they are strongly counter-cyclical --- this highlights the fact that the estimated structural shock behaves like a financial stock. The only noticeable difference is that our median responses of prices are all positive, whereas those in \citet{FRS19} have mixed signs (although the credible intervals in both cases are relatively wide and include 0 for some periods).

As a comparison, we also compute the corresponding impulse responses under a standard independent normal and inverse-Wishart priors: a Minnesota prior on the VAR coefficients with hyperparameters $\tilde{\kappa}_1 = \tilde{\kappa}_2 = 1$ and an inverse-Wishart prior on $\tilde{\vSigma}$. The results are almost identical to those obtained under the asymmetric conjugate prior (results are reported in Appendix E), suggesting any minor differences from \citet{FRS19} are likely due to a slightly different dataset used.

\begin{figure}[H]
    \centering
   \includegraphics[width=1\textwidth]{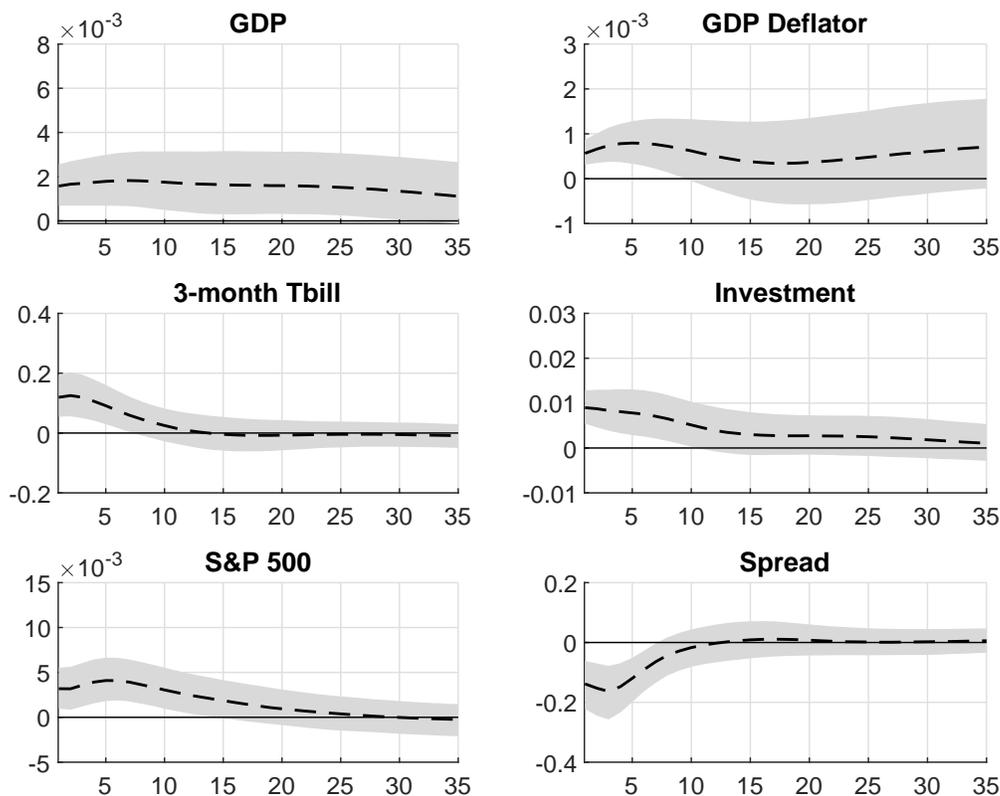}
   \caption{Impulse responses from a 15-variable VAR with the asymmetric conjugate prior to an one-standard-deviation financial shock. The shaded region represents the 16-th and 84-th percentiles.}
   \label{fig:IRF_finc_15}	
\end{figure}

Next, we construct the impulse responses from the 15-variable VAR. Given the large number of VAR coefficients, appropriate shrinkage is vital in this case. We therefore obtain the optimal shrinkage hyperparameters by maximizing the marginal likelihood of the VAR with the asymmetric conjugate prior. As reported in Section~\ref{ss:fullsample}, the optimal shrinkage hyperparameters are found to be 
$\tilde{\kappa}_1 = 0.058$ and $\tilde{\kappa}_2 = 0.0043$. The following results are based on these hyperparameter values.

Figure~\ref{fig:IRF_finc_15} reports the impulse responses of the same 6 variables to an one-standard-deviation financial shock. Compared to the results from the 6-variable VAR depicted in Figure~\ref{fig:IRF_finc_6}, the median responses are mostly the same, but the credible intervals are substantially narrower. For example, the responses of GDP are much more precisely estimated --- the credible intervals exclude 0 for the first~30 quarters after impact. These results highlight the usefulness of the proposed asymmetric conjugate prior and the value of including more variables and sign restrictions in identifying structural shocks.

\begin{figure}[H]
    \centering
   \includegraphics[width=1\textwidth]{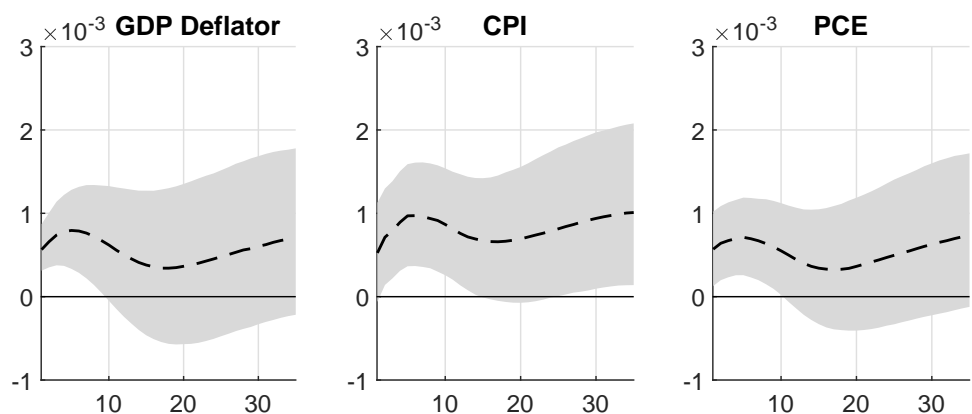}
   \caption{Impulse responses of prices to an one-standard-deviation financial shock. The shaded region represents the 16-th and 84-th percentiles.}
   \label{fig:IRF_finc_prices}	
\end{figure}

As mentioned earlier, there are typically multiple data series corresponding to the same economic variable. And it is often unclear which one should be used, if only one variable is to be selected. In our application, for example, the time series GDP deflator, CPI and PCE are all good candidates for the economic variable prices. Instead of picking one out of the three candidates, we include them all in the analysis. Figure~\ref{fig:IRF_finc_prices} reports impulse responses of the three measure of prices to an one-standard-deviation financial shock. The median responses of the three variables are all positive and have very similar shapes. However, their credible intervals are somewhat different --- only in the case of CPI do the credible intervals mostly exclude 0. If one had only used CPI, one might have drawn a stronger conclusion than it is warranted. This again highlights the benefit of including more variables and performing a more comprehensive structural analysis.

\section{Concluding Remarks and Future Research} \label{s:conclusion}

We developed a new asymmetric conjugate prior for large BVARs that can accommodate cross-variable shrinkage, while maintaining many useful analytical results as the natural conjugate prior. Using the new prior and the associated efficient sampler, we were able to identify 5 structural shocks using a 15-variable VAR with a set of  sign restrictions on the contemporaneous impact matrix. We showed that the larger number of variables and sign restrictions --- when used in conjunction with the asymmetric conjugate prior --- can sharpen inference on impulse responses.
	

There is now a large empirical literature that shows that models with stochastic volatility tend to forecast substantially better \citep{clark11, DGG13, CP16}. 
In future work, it would be useful to develop similar efficient posterior samplers for large BVARs with stochastic volatility, such as the models in \citet*{CCM19} and \citet*{CES20}.  In addition, it would be interesting to use Proposition 1 to construct a multivariate stochastic volatility model that is invariant to reordering of the variables.

\newpage

\section*{Appendix A: Data}

This appendix provides the details of the raw data used to construct the variables in the empirical application. In particular, Table~\ref{tab:var} lists the variables and their sources. The sample period is from 1985:Q1 to 2019:Q4.

\begin{table}[H]
\centering
\caption{Description of variables used in the empirical application.} \label{tab:var}
\resizebox{\textwidth}{!}{\begin{tabular}{lll}
\hline\hline
Variable 	&	Description	&	Source	\\ \hline
GDP	&	Log of real GNP/GDP	&	Federal Reserve Bank of Philadelphia	\\ 
\rowcolor{lightgray}
GDP Deflator	&	Log of price index of GNP/GDP 	&	Federal Reserve Bank of Philadelphia	\\
3-month treasury bill	&	3-month treasury bill rate	&	Federal Reserve Bank of St. Louis	\\
\rowcolor{lightgray}
Investment	&	Log of real gross private domestic investment	&	Federal Reserve Bank of St. Louis	\\
S\&P 500	&	Log of S\&P 500	&	Yahoo Finance	\\
\rowcolor{lightgray}
Total credit	&	Log of loans to non-financial private sector	&	Board of Governors of the Federal \\
\rowcolor{lightgray}
& & Reserve System	\\
Mortgages	&	Log of home mortgages of households and & Federal Reserve Bank of St. Louis \\
  & non-profit organizations	&		\\
\rowcolor{lightgray}
Real estate value	&	Log of real estate at market value of households & Federal Reserve Bank of St. Louis \\ 
\rowcolor{lightgray}
 & and non-profit organizations	&		\\
Corporate bond yield	&	Moody's baa corporate bond yield	&	Federal Reserve Bank of St. Louis	\\
\rowcolor{lightgray}
10-year treasury note	&	10-year treasury constant maturity rate	&	Federal Reserve Bank of St. Louis	\\
Federal funds rate	&	Federal funds rate	&	Federal Reserve Bank of St. Louis	\\
\rowcolor{lightgray}
Mortgage rate	&	30-year fixed rate mortgage average	&	Federal Reserve Bank of St. Louis	\\
CPI	&	Log of consumer price index	&	Federal Reserve Bank of St. Louis	\\
\rowcolor{lightgray}
PCE	&	Log of price index of personal consumption &	Federal Reserve Bank of St. Louis	\\
\rowcolor{lightgray}
& expenditure & \\
Employment	&	Log of employment level	&	Federal Reserve Bank of St. Louis	\\
\rowcolor{lightgray}
Industrial production	&	Log of industrial production index	&	Federal Reserve Bank of St. Louis	\\
1-year treasury bill	&	1-year treasury constant maturity rate	&	Federal Reserve Bank of St. Louis	\\
\rowcolor{lightgray}
Dow Jones Industrial Average	&	Log of Dow Jones Industrial Average index	&	Google Finance	\\
\hline\hline
\end{tabular}
}
\end{table}

\newpage

\section*{Appendix B: Derivation of the Marginal Likelihood}

In this appendix we prove that the marginal likelihood of the VAR($p$) under the asymmetric conjugate prior in \eqref{eq:ACP} has the following expression:
\[
	p(\by) = \prod_{i=1}^n (2\pi)^{-\frac{T}{2}} |\bV_i|^{-\frac{1}{2}} |\bK_{\vtheta_i}|^{-\frac{1}{2}} \frac{\Gamma(\nu_i+T/2) S_i^{\nu_i}}{\Gamma(\nu_i) \hat{S}_i^{\nu_i+\frac{T}{2}}}.
\]
This result follows from direct computation:
\begin{align*}
	p(\by)  = \prod_{i=1}^n p(\by_i) & = \prod_{i=1}^n	\int p(\vtheta_i,\sigma^2_i)p(\by_i\gvn\vtheta_i,\sigma^2) \di(\vtheta_i,\sigma^2_i)\\
	& = \prod_{i=1}^n c_i(2\pi)^{-\frac{T}{2}}
	\int  (\sigma^2_i)^{-\left(\nu_i+\frac{T+k_i}{2}+1\right)}
			\e^{-\frac{1}{\sigma^2_i}\left(\hat{S}_i + \frac{1}{2}(\vtheta_i-\hat{\vtheta}_i)'\bK_{\vtheta_i}(\vtheta_i-\hat{\vtheta}_i)\right)}
	 \di(\vtheta_i,\sigma^2_i)\\
	& = \prod_{i=1}^n c_i(2\pi)^{-\frac{T}{2}}  (2\pi)^{\frac{k_i}{2}} |\bK_{\vtheta_i}^{-1}|^{\frac{1}{2}}\frac{\Gamma(\nu_i+T/2)}{\hat{S}_i^{\nu_i+\frac{T}{2}}} \\
	& = \prod_{i=1}^n (2\pi)^{-\frac{T}{2}} |\bV_i|^{-\frac{1}{2}} |\bK_{\vtheta_i}|^{-\frac{1}{2}} \frac{\Gamma(\nu_i+T/2) S_i^{\nu_i}}{\Gamma(\nu_i) \hat{S}_i^{\nu_i+\frac{T}{2}}}.
\end{align*}
where $c_i = (2\pi)^{-\frac{k_i}{2}}|\bV_i|^{-\frac{1}{2}}S_i^{\nu_i}/\Gamma(\nu_i)$, $\bK_{\vtheta_i} = \bV_i^{-1} + \bX_i'\bX_i$, $\hat{\vtheta}_i = \bK_{\vtheta_i}^{-1}(\bV_i^{-1}\bm_i+ \bX_i'\by_i)$ and $\hat{S}_i = S_i + (\by_i'\by_i + \bm_i'\bV_i^{-1}\bm_i - \hat{\vtheta}_i'\bK_{\vtheta_i}\hat{\vtheta}_i)/2$. In the above derivation we have used the fact that 
\[
\int  (\sigma^2_i)^{-\left(\nu_i+\frac{T+k_i}{2}+1\right)} \e^{-\frac{1}{\sigma^2_i}\left(\hat{S}_i + \frac{1}{2}(\vtheta_i-\hat{\vtheta}_i)'\bK_{\vtheta_i}(\vtheta_i-\hat{\vtheta}_i)\right)} 
	 =  (2\pi)^{\frac{k_i}{2}} |\bK_{\vtheta_i}^{-1}|^{\frac{1}{2}}\frac{\Gamma(\nu_i+T/2)}{\hat{S}_i^{\nu_i+\frac{T}{2}}}.
\]
The above equality holds because the quantity on the right-hand side is the normalizing constant of the  $(\vtheta_i,\sigma^2_i)\sim\distn{NIG}(\bm_{i},\bV_{i},\nu_{i}, S_{i})$ distribution.

\newpage 

\section*{Appendix C: Proofs of Proposition and Corollaries}

This appendix provides proofs of Proposition~\ref{thm:main} and two corollaries in the main text. We first record in the following lemma the determinant of the Jacobian of transformation from the structural-form parameterization to the reduced-form parameterization.\footnote{For a different structural-form parameterization where the error covariance matrix is the identity matrix and the impact matrix is triangular with free diagonal elements, \citet{zha99} derives the determinant of the Jacobian of transformation from the reduced-form parameterization.} This lemma was proved in \citet{CJ09b}, and we include it here for convenience. The proof uses the differential forms approach that is equivalent to calculating the Jacobian \citep[see, e.g., Theorem 2.1.1 in][]{muirhead}. 

\begin{lemma}\label{thm:lem1} \rm Suppose $\bW$ is a $n\times n$ positive definite matrix and $\bW=\bT'\tilde{\bT}\bT$, where $\bT$ is a lower triangular matrix with ones on the main diagonal
and $\tilde{\bT}$ is a diagonal matrix with positive diagonal elements. Denote the lower diagonal elements of $\bT$ by $t_{ij},\; 1\leq j<i\leq n,$ and the 
diagonal elements of $\tilde{\bT}$ by $t_{ii},\; i=1,\ldots, n,$. Let $(\di\bW)$ denote the differential form $(\di\bW)\equiv \mathop{\wedge}\limits_{i \geq j}\di w_{ij}$ and similarly define 
$(\di \bT) \equiv \mathop{\wedge}\limits_{i \geq j}\di t_{ij}$. Then we
have
\[
    (\di \bW) = \prod_{i=1}^{n}t_{ii}^{i-1}(\di\bT).
\]
In other words, the determinant of the Jacobian of the transformation from $\bT'\tilde{\bT}\bT$ to $\bW$ is $\prod_{i=1}^{n}t_{ii}^{-i+1}$.
\end{lemma}

\textbf{Proof of the lemma}: By the definition $\bW=\bT'\tilde{\bT}\bT$, we have
\[
	\begin{pmatrix}        
        w_{11}  & w_{21}& \ldots & w_{n1} \\
        w_{21}  & w_{22}& \ldots & \vdots \\
        \vdots  & \vdots& \ddots &\vdots \\
        w_{n1}  & w_{n2}& \ldots & w_{nn}
	\end{pmatrix}
     =
  \begin{pmatrix}   
        1   & t_{21}& \ldots & t_{n1} \\
        0   & 1 & \ldots & t_{n2} \\
        \vdots  & \vdots& \ddots & \vdots \\
        0   & 0 & \ldots & 1
  \end{pmatrix}
  \begin{pmatrix}   
        t_{11}  & 0 & \ldots& 0 \\
        0   & t_{22}& \ldots& 0 \\
        \vdots  & \vdots & \ddots   & \vdots \\
        0   & 0  & \ldots   & t_{nn}
  \end{pmatrix}
  \begin{pmatrix}   
        1   & 0 & \ldots & 0 \\
        t_{21}  & 1 & \ldots & 0 \\
        \vdots  & \vdots& \ddots & \vdots \\
        t_{n1}  & t_{n2} & \ldots & 1
	\end{pmatrix}.
\]
Hence, we can express each $w_{ij}$ in terms of $\{ t_{ij}\}$:
\begin{align}
    w_{ii}  & = t_{ii} + \sum_{j=i+1}^{n}t_{ji}^{2}t_{jj}, \quad i=1,\ldots, n, \label{eq:wii} \\
    w_{ij}  & = t_{ij}t_{ii} + \sum_{k=i+1}^{n}t_{ki}t_{kj}t_{kk}, \quad 1 \leq j<i\leq n. \label{eq:wij}
\end{align}
Next, we take differentials of these two equations so that we can write the differential form  $(\di\bW)$ in terms of  $(\di\bT)$.
Since we are going to take the exterior product of these differentials and the exterior products of repeated differentials are zero, there is no need to keep track
of differentials in $t_{ij}$ that have previously occurred. Therefore, we take differentials of \eqref{eq:wii} and \eqref{eq:wij} and ignore those that have previously occurred:
\begin{align*}
    \di w_{nn} & = \di t_{nn} \\
    \di w_{n,n-1} & = \di t_{nn}\di t_{n,n-1} + \ldots \\
    & \vdots \\
    \di w_{n1} & =  t_{nn}\di t_{n1} + \ldots \\
    \di w_{n-1,n-1} & =  \di t_{n-1,n-1} + \ldots \\
    & \vdots \\
    \di w_{11} & =  \di t_{11} +\ldots
\end{align*}
Finally, taking exterior products gives
\begin{equation*}
    \mathop{\wedge}\limits_{i \geq j}\di w_{ij} = t_{nn}^{n-1}t_{n-1,n-1}^{n-2}\cdots t_{22}\mathop{\wedge}\limits_{i \geq j}\di t_{ij}
\end{equation*}
as claimed.  \hfill $\square$

\textbf{Proof of Proposition~\ref{thm:main}}: Assume the same notation as in Lemma~\ref{thm:lem1}. To prove Proposition~\ref{thm:main}, we consider the case where 
\begin{align}
	t_{ii} & \sim \distn{G}\left(\frac{\nu_0+i-n}{2},\frac{s_i^2}{2}\right), \quad i=1,\ldots, n, \label{eq:tii} \\	
	(t_{ij}\gvn t_{ii}) & \sim \distn{N}\left(0, \frac{t_{ii}^{-1}}{s_j^2}\right), \quad 1\leq j<i\leq n, \; i=2,\ldots, n. \label{eq:tij}
\end{align}
More specifically, we will show that the density of $\bW=\bT'\tilde{\bT}\bT$ is the same as that of the Wishart distribution $\distn{W}(\nu, \bS^{-1})$, where $ \bS = \text{diag}(s_1^2,\ldots, s_n^2)$.
Then, if we let $t_{ii} = 1/\sigma_i^{2}$ and $A_{i,j} = t_{ij}$, we have $\tilde{\vSigma}^{-1} = \bA'\vSigma^{-1}\bA \sim \distn{W}(\nu_0, \bS^{-1})$.  

To prove the proposition, we first compute the determinant of $\bW$ and the trace $\text{tr}(\bS \bW)$. Since the determinant of $\bT$ is 1, we have
\[
    |\bW| = |\tilde{\bT}| = \prod_{i=1}^{n}t_{ii}.
\]
Next, using \eqref{eq:wii}, we have
\begin{align*}
    \text{tr}(\bS \bW) & = \sum_{i=1}^{n }w_{ii} s_i^2 \\
        & = \sum_{i=1}^{n}t_{ii}s_i^2 + \sum_{i=1}^{n}\sum_{j=i+1}^{n}t_{ji}^{2}t_{jj}s_i^2 \\
        & = \sum_{i=1}^{n}t_{ii}s_i^2 + \sum_{j=2}^{n}\sum_{i=1}^{j-1}t_{ji}^{2}t_{jj}s_i^2 \\
        & = \sum_{i=1}^{n}t_{ii}s_i^2 + \sum_{i=2}^{n}\sum_{j=1}^{i-1}t_{ij}^{2}t_{ii}s_j^2,
\end{align*}
where we change the order of the double summations in the third equality and interchange the dummy indices $i$ and $j$ in the last equality. 

Now, it follows from the distributional assumptions in~\eqref{eq:tii} and~\eqref{eq:tij} that the kernel of the joint density of $\bT$ and $\tilde{\bT}$ is
\begin{align*}
    & \prod_{i=1}^{n}t_{ii}^{\frac{\nu_0+i-n}{2}-1}\e^{-\frac{s_i^2}{2}t_{ii}}\times \prod_{i=2}^{n}t_{ii}^{\frac{i-1}{2}}\e^{-\frac{1}{2}\sum_{j=1}^{i-1}t_{ij}^{2}t_{ii}s_j^2} \\
    & = \left(\prod_{i=1}^{n}t_{ii}^{\frac{\nu_0-n-1}{2}+(i-1)}\right)\e^{-\frac{1}{2} \left( \sum_{i=1}^{n}t_{ii}s_i^2 + \sum_{i=2}^{n}\sum_{j=1}^{i-1}t_{ij}^{2}t_{ii}s_j^2 \right)}.
\end{align*}
Next, we derive the kernel of the density of $\bW$. By the lemma, the determinant of the Jacobian is 
$\prod_{i=1}^{n}t_{ii}^{-i+1}$. Substituting $\text{tr}(\bW)$ and $|\bW|$ into the above expression and multiplying the determinant of the Jacobian, 
we obtain the kernel of the density of $\bW$:
\[
    |\bW|^{\frac{\nu_0-n-1}{2}}\e^{-\frac{1}{2}\text{tr}(\bS\bW)},
\]
which is the kernel of the Wishart density $\mathcal{W}(\nu_0, \bS^{-1})$. $\hfill \square$

\textbf{Proof of Corollary~\ref{thm:cor1}}: Here we use the same notation as in Proposition 1. 
Assume $\bW\sim \distn{W}(\nu, \bS^{-1})$, and let $\bW=\bT'\tilde{\bT}\bT$, where $\bT$ and 
$\tilde{\bT}$ are given in Lemma~\ref{thm:lem1}. If we can show that $t_{ii}$ and 
$(t_{ij}\gvn t_{ii})$ follow the same normal-gamma distributions given in \eqref{eq:tii} and \eqref{eq:tij}, respectively, then we are done.
Since the transformation between $\bW$ and $\bT'\tilde{\bT}\bT$ is one-to-one, the proof essentially just 
``reverses" the equalities given in Proposition 1. More specifically, in the proof of Proposition 1 we showed that $|\bW| = \prod_{i=1}^{n}t_{ii}$ and 
\[
    \text{tr}(\bS \bW) = 
		\sum_{i=1}^{n}t_{ii}s_i^2 + \sum_{i=2}^{n}\sum_{j=1}^{i-1}t_{ij}^{2}t_{ii}s_j^2.
\]
Also, by Lemma~1, the determinant of the Jacobian of transformation is $\prod_{i=1}^{n}t_{ii}^{i-1}$. Hence, the kernel of the joint distribution of $t_{ii}, i=1,\ldots, n,$ and $t_{ij}, 1\leq j<i\leq n, \; i=2,\ldots, n$ is given by:
\begin{align*}
	  |\bW|^{\frac{\nu_0-n-1}{2}}&\e^{-\frac{1}{2}\text{tr}(\bS\bW)}\times \prod_{i=1}^{n}t_{ii}^{i-1} \\
		= & \left(\prod_{i=1}^{n}t_{ii}^{\frac{\nu_0-n-1}{2}+(i-1)}\right)\e^{-\frac{1}{2} \left( \sum_{i=1}^{n}t_{ii}s_i^2 + \sum_{i=2}^{n}\sum_{j=1}^{i-1}t_{ij}^{2}t_{ii}s_j^2 \right)} \\
		= & \prod_{i=1}^{n}t_{ii}^{\frac{\nu_0+i-n}{2}-1}\e^{-\frac{s_i^2}{2}t_{ii}}\times \prod_{i=2}^{n}t_{ii}^{\frac{i-1}{2}}\e^{-\frac{1}{2}\sum_{j=1}^{i-1}t_{ij}^{2}t_{ii}s_j^2}.
\end{align*}
It follows that $t_{ii}, i=1,\ldots, n,$ are independent gamma random variables given in \eqref{eq:tii}.
Moreover, conditional on $t_{ii}$, $t_{ij},1\leq j<i, $ are independent normal variables given in~\eqref{eq:tij}.

\textbf{Proof of Corollary~\ref{thm:cor3}}: Suppose $\tilde{\vSigma} \sim \distn{IW}(\nu_0, \bR)$, where $\bR$ is a symmetric positive definite matrix. Factor $\bR^{-1}  = \bL' \bS^{-1} \bL$, where 
$\bL$ is lower triangular with ones on the main diagonal and $\bS$ is diagonal. Since $\tilde{\vSigma}^{-1} \sim \distn{W}(\nu_0, \bR^{-1})$, by the properties of the Wishart distribution, 
we have $(\bL')^{-1}\tilde{\vSigma}^{-1}\bL^{-1} \sim \distn{W}(\nu_0, \bS^{-1})$. Now, applying Corollary~1, we obtain $(\bL')^{-1}\tilde{\vSigma}^{-1}\bL^{-1} = \bA' \vSigma^{-1} \bA$, where $\bA$ is lower triangular with ones on the main diagonal and $\vSigma$ is diagonal. The diagonal elements of $\vSigma$ and the lower triangular elements of $\bA$ follow the normal-inverse-gamma distributions:
\begin{align*}
	\sigma_i^{2} & \sim \distn{IG}\left(\frac{\nu_0+i-n}{2},\frac{s_i^2}{2}\right), \quad i=1,\ldots, n,\\	
	(A_{i,j}\gvn \sigma_i^{2}) & \sim \distn{N}\left(0, \frac{\sigma_i^2}{s_j^2}\right), \quad 1\leq j<i\leq n, \; i=2,\ldots, n.
\end{align*}
Letting $\bC = \bA\bL$, we can write $\tilde{\vSigma}^{-1} = \bC' \vSigma^{-1} \bC$. Since both $\bA$ and $\bL$ are lower triangular with ones on the main diagonal, so is $\bC$.
It remains to show that $\bc_i$, the free elements of the $i$-th row of $\bC$, follows the normal distribution in \eqref{eq:prior2_Cij}. Since  $\bC' = \bL'\bA'$, we can 
write $\bc_i$ in terms of $\bA$ and $\bL$ as:
\[
	\bc_i = \mathbf{l}_i + \bL_{1:i-1}' \mathbf{a}_i,
\]
where $\mathbf{l}_i$ and $\mathbf{a}_i$ are respectively the free elements of the $i$-th row of $\bL$ and $\bA$, and $ \bL_{1:i-1}$ is the $(i-1)\times(i-1)$ matrix that consists of the first $(i-1)$ rows and columns of $\bL$. Since $\bc_i$ is an affine transformation of the normal vector $\mathbf{a}_i$, conditional on $\sigma_i^2$, $\bc_i$ is normally distributed with mean vector $\mathbf{l}_i$ and covariance matrix  $\sigma_i^2\bL_{1:i-1}'\bS_{1:i-1}^{-1}\bL_{1:i-1}$, where $\bS_{1:i-1}$ is the submatrix consisting of the first $(i-1)$ rows and columns of 
$\bS = \text{diag}(s_1^2,\ldots, s_n^2)$. $ \mbox{} \hfill \square$

\newpage

\section*{Appendix D: Derivation of the Implied Structural-Form Hyperparameters}

This appendix derives the implied prior hyperparameters of the structural-form VAR coefficients from the hyperparameters of the reduced-form VAR coefficients. Let $(B_k)_{ij}$ and  
$(\tilde{B}_{k})_{ij}$ denote the $(i,j)$-th element of the structural-form coefficient matrix $\bB_k$ and the reduced-form coefficient matrix $\tilde{\bB}_k$, respectively, for $i,j=1,\ldots, n, k = 1,\ldots, p$. 
Further, let $\bm_{\tilde{\vbeta},i}$ and  $\sigma_{i}^2\bV_{\tilde{\vbeta},i}$ denote the prior mean vector and covariance matrix of the reduced-form parameters 
$\tilde{\vbeta}_i = (\tilde{b}_i,(\tilde{B}_1)_{i1},\ldots,(\tilde{B}_1)_{in},\ldots,(\tilde{B}_p)_{i1},\ldots,(\tilde{B}_p)_{in})'$. The hyperparameters $\bm_{\tilde{\vbeta},i}$ and 
$\bV_{\tilde{\vbeta},i}$ can be elicited in a standard way --- e.g., following the Minnesota prior in \citet*{DLS84} and \citet{litterman86}. We further assume that $\bV_{\tilde{\vbeta},i}$ is diagonal as is usually done in the literature. The goal is to derive the structural-form hyperparameters $\bm_{\vbeta,i}$ and  $\bV_{\vbeta,i}$ given 
$\bm_{\tilde{\vbeta},i}$ and  $\bV_{\tilde{\vbeta},i}$.

To that end, recall that the structural-form and reduced-form coefficients are related via
\[
	(B_k)_{ij} = (\tilde{B}_k)_{ij} + \sum_{l=1}^{i-1} A_{i,l} (\tilde{B}_k)_{lj}.
\]
Using the law of iterated expectations and the assumption that the prior means of $\valpha_i = (A_{i,1}, \ldots, A_{i,i-1})'$ are zero (see Proposition~\ref{thm:main}), we obtain
\[
	\Em (B_k)_{ij} = \Em\left[\Em((B_k)_{ij} \gvn  \tilde{\bB}_{k})\right] = \Em (\tilde{B}_k)_{ij}.
\]
Hence, we have $\bm_{\vbeta,i} = \bm_{\tilde{\vbeta},i}$.

Next, we elicit the prior covariance matrix $\bV_{\vbeta,i}$, which is in general a full matrix. Here for simplicity we ignore the correlations between the structural-form coefficients and set
$\bV_{\vbeta,i}$ to be diagonal. One could work out all the covariances following a similar derivation presented below. Now, we derive the variance of a generic element $(B_k)_{ij}$. Let $f(j,k) = (k-1)n + j+ 1$ be an integer-value function keeping track of the indices. Further, let 
$(V_{\tilde{\vbeta},i})_{f(j,k)} $ and $(m_{\tilde{\vbeta},i})_{f(j,k)}$ denote the $f(j,k)$-th diagonal element of $\bV_{\tilde{\vbeta},i}$ and the $f(j,k)$-th element of $\bm_{\tilde{\vbeta},i}$, respectively. Then, we have 
\begin{align*}
	\text{Var}((B_k)_{ij}) & = \Em\left[\text{Var}((B_k)_{ij}\gvn \bA)  \right] + \text{Var}\left[\Em( (B_k)_{ij}\gvn \bA)\right] \\
	 & = \Em\left[ \text{Var}((\tilde{B}_k)_{ij}) + \sum_{l=1}^{i-1} A_{i,l}^2\text{Var}((\tilde{B}_k)_{lj})\right] 
			+ \text{Var}\left[  \Em(\tilde{B}_k)_{ij} + \sum_{l=1}^{i-1} A_{i,l}\Em(\tilde{B}_k)_{lj} \right] \\
	& = \text{Var}((\tilde{B}_k)_{ij}) + \sum_{l=1}^{i-1} \text{Var}(A_{i,l})\text{Var}((\tilde{B}_k)_{lj}) + \sum_{l=1}^{i-1} \text{Var}(A_{i,l})(\Em(\tilde{B}_k)_{lj})^2 \\
	& = \sigma_i^2 (V_{\tilde{\vbeta},i})_{f(j,k)} + \sum_{l=1}^{i-1}\frac{\sigma_i^2}{s_l^2}\left[\sigma_l^2 (V_{\tilde{\vbeta},l})_{f(j,k)} + (m_{\tilde{\vbeta},l})_{f(j,k)}^2\right].
\end{align*}
Note that this variance involves the error variances of other equations, namely, $\sigma_1^2, \ldots, \sigma_{i-1}^2$. To avoid this issue, we replace $\sigma_l^2$ by its prior mean $s_l^2$. Using this simplification, we obtain:
\[
	\text{Var}((B_k)_{ij}) \approx \sigma_i^2\left[ (V_{\tilde{\vbeta},i})_{f(j,k)} + \sum_{l=1}^{i-1}\left( (V_{\tilde{\vbeta},l})_{f(j,k)} + s_l^{-2}(m_{\tilde{\vbeta},l})_{f(j,k)}^2\right)\right].
\]
Given this approximation, we set the diagonal element in $\bV_{\vbeta,i}$ associated with $(B_k)_{ij}$ to be 
$(V_{\tilde{\vbeta},i})_{f(j,k)} + \sum_{l=1}^{i-1}( (V_{\tilde{\vbeta},l})_{f(j,k)} + s_l^{-2}(m_{\tilde{\vbeta},l})_{f(j,k)}^2)$.

\newpage

\section*{Appendix E: Comparison with Independent Normal and Inverse-Wishart Prior}

This appendix reports impulse responses of the 6-variable VAR described in Section~\ref{s:application} under the independent normal and inverse-Wishart priors. More specifically, we consider a standard Minnesota prior on the VAR coefficients with hyperparameters $\tilde{\kappa}_1 = \tilde{\kappa}_2 = 1$ and an inverse-Wishart prior on the reduced-form error covariance matrix $\tilde{\vSigma}\sim \distn{IW}(\nu_0,\mathbf{S})$. Given the posterior draws of the structural-form parameters, we transform them to the corresponding reduced-form parameters. We then use the algorithm described in \citet{RWZ10} to incorporate the sign restrictions to construct impulse responses. 

\begin{figure}[H]
    \centering
   \includegraphics[width=1\textwidth]{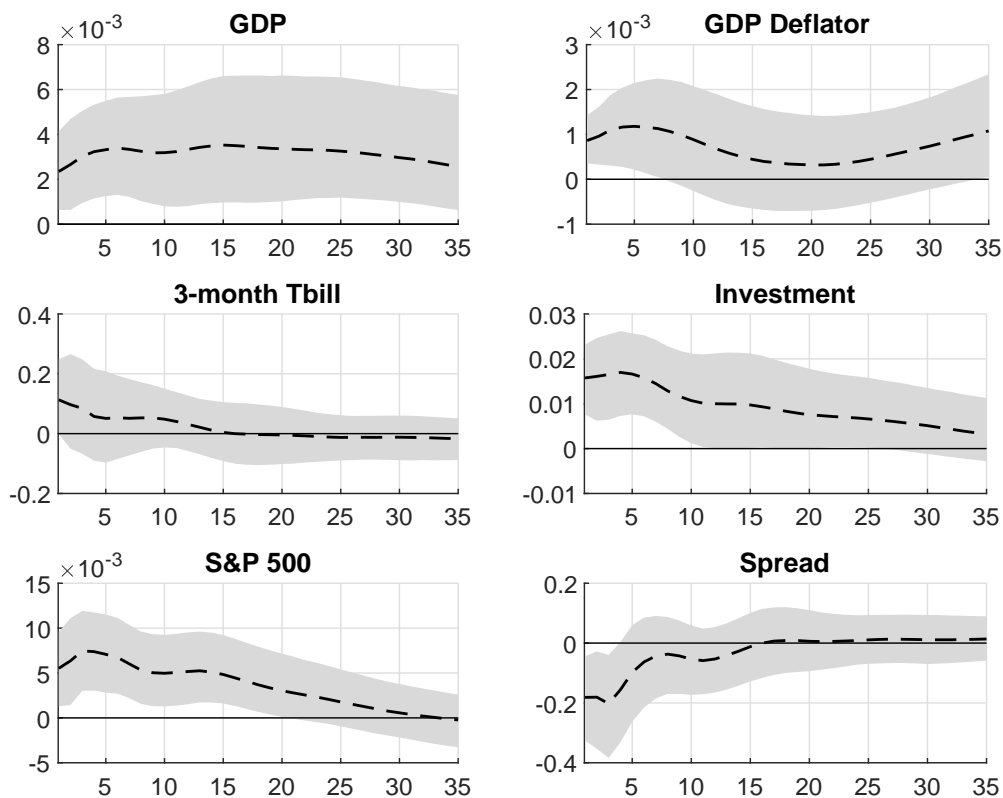}
   \caption{Impulse responses from a 6-variable VAR with the indpendent normal and inverse-Wishart priors to an one-standard-deviation financial shock. The shaded region represents the 16-th and 84-th percentiles.}
   \label{fig:IRF_finc_6_IP}	
\end{figure}

Figure~\ref{fig:IRF_finc_6_IP} reports the impulse responses to an one-standard-deviation financial shock. These results are almost identical to those obtained under the asymmetric conjugate prior with the same  hyperparameters $\tilde{\kappa}_1 = \tilde{\kappa}_2 = 1$ reported in Figure~\ref{fig:IRF_finc_6} in the main text.

\newpage 

\singlespace
\ifx\undefined\BySame
\newcommand{\BySame}{\leavevmode\rule[.5ex]{3em}{.5pt}\ }
\fi
\ifx\undefined\textsc
\newcommand{\textsc}[1]{{\sc #1}}
\newcommand{\emph}[1]{{\em #1\/}}
\let\tmpsmall\small
\renewcommand{\small}{\tmpsmall\sc}
\fi

\end{document}